\documentclass[twocolumn]{aastex62}

\usepackage{gensymb,amsmath}
\usepackage{epstopdf}
\usepackage{url}
\usepackage[]{natbib}

\usepackage{bm}

\usepackage{float}
\usepackage{subfigure}

\usepackage{multirow}

\newcommand{\ci}{C\,{\sc i }(${\rm ^3P_2\rightarrow ^3\!P_1}$) }
\newcommand{\cishort}{C\,{\sc i }}

\shorttitle{Resolved SFR and Gas Tracers in J0901}
\shortauthors{Chen et al.}

\begin{document}

\title{Comparisons Between Resolved Star Formation Rate and Gas Tracers in the Strongly Lensed Galaxy SDSS\,J0901+1814 at Cosmic Noon}

\correspondingauthor{Qingxiang Chen}
\email{chenqingxiangcn@gmail.com}

\author[0000-0002-3864-0965]{Qingxiang Chen}
\affil{Yale-NUS College, 16 College Avenue West \#01-220, Singapore, 138527, Singapore}

\author{Chelsea E. Sharon}
\affil{Yale-NUS College, 16 College Avenue West \#01-220, Singapore, 138527, Singapore}

\author{Hiddo S. B. Algera}
\affil{Hiroshima University, 1-3-2, Kagamiyama, Higashi-hiroshima, Hiroshima 739-0046, Japan}
\affil{National Astronomical Observatory of Japan, 2-21-1, Osawa, Mitaka, Tokyo, Japan}

\author{Andrew J. Baker}
\affil{Department of Physics and Astronomy, Rutgers, The State University of New Jersey, 136 Frelinghuysen Road, Piscataway, NJ 08854-8019, USA}
\affil{Department of Physics and Astronomy, University of the Western Cape, Robert Sobukwe Road, Bellville 7535, South Africa}

\author{Charles R. Keeton}
\affil{Department of Physics and Astronomy, Rutgers, The State University of New Jersey, 136 Frelinghuysen Road, Piscataway, NJ 08854-8019, USA}

\author{Dieter Lutz}
\affil{Max-Planck-Institut f{\"u}r extraterrestrische Physik}

\author[0000-0001-9773-7479]{Daizhong Liu}
\affiliation{Max-Planck-Institut f\"ur Extraterrestrische Physik (MPE), Giessenbachstr. 1, D-85748 Garching, Germany}

\author{Anthony J. Young}
\affil{Department of Physics and Astronomy, Rutgers, The State University of New Jersey, 136 Frelinghuysen Road, Piscataway, NJ 08854-8019, USA}

\author{Amitpal S. Tagore}
\affil{Jodrell Bank Centre for Astrophysics, The University of Manchester, Manchester, M13 9PL, UK}

\author{Jesus Rivera}
\affil{Swarthmore College, Department of Physics and Astronomy, Swarthmore, PA 19081}

\author{Erin K. S. Hicks}
\affil{Department of Physics and Astronomy, University of Alaska, Anchorage, AK 99508, USA}

\author{Sahar S. Allam}
\affil{Center for Particle Astrophysics, Fermi National Accelerator Laboratory, P.O. Box 500, Batavia, IL 60510, USA}

\author{Douglas L. Tucker}
\affil{Center for Particle Astrophysics, Fermi National Accelerator Laboratory, P.O. Box 500, Batavia, IL 60510, USA}

\defcitealias{Sharon:2019a}{S19}

\begin{abstract}

We report new radio observations of SDSS\,J090122.37+181432.3, a strongly lensed star-forming galaxy at $z=2.26$. We image 1.4\,GHz (L-band) and 3\,GHz (S-band) continuum using the VLA and 1.2\,mm (band 6) continuum with ALMA, in addition to the CO(7--6) and \ci lines, all at $\lesssim1.^{\prime\prime}7$ resolution. Based on the VLA integrated {flux densities}, we decompose the radio spectrum into its free-free (FF) and non-thermal components. The {infrared-radio correlation (IRRC) parameter $q_{\rm TIR}=2.65_{-0.31}^{+0.24}$ is consistent with expectations for star forming galaxies}. We obtain radio continuum-derived SFRs that are free of dust extinction, finding {${\rm{620}_{-220}^{+280}\,M_\odot\,yr^{-1}}$, ${\rm {230}_{-160}^{+570}\,M_\odot\,yr^{-1}}$, and ${\rm {280}_{-120}^{+460}\,M_\odot\,yr^{-1}}$ from the} FF emission, non-thermal emission{, and when accounting for both emission processes,} respectively, in agreement with previous results. We estimate the gas mass from the \ci line as $M_{\rm gas}=(1.2\pm0.2)\times10^{11}\,M_\sun$, which is consistent with prior CO(1--0)-derived gas masses. Using our new IR and radio continuum data to map the SFR, we assess the dependence of the Schmidt-Kennicutt relation on choices of SFR and gas tracer {for $\sim{\rm kpc}$ scales}. The different SFR tracers yield different slopes, with the IR being the steepest, potentially due to highly obscured star formation in J0901. The radio continuum maps have the lowest slopes and overall fidelity for mapping the SFR, despite producing consistent total SFRs. We also find that the Schmidt-Kennicutt relation slope is flattest when using CO(7--6) or \ci to trace gas mass, suggesting that those transitions are not suitable for tracing the bulk molecular gas in galaxies like J0901.

\end{abstract}

\section{Introduction}\label{sec:intro}

How galaxies form and evolve remains an essential question yet to be fully understood. A simple `bathtub' model proposes a close connection between galaxies' star formation activity and the gas supply in the local environment \citep{Lilly:2013a}. Gas is consumed by star formation and will also be removed by stellar winds and supernova explosions. Therefore, in addition to studying gas inflows and outflows on halo scales, understanding the full picture of galaxy evolution requires accurate measurements of galaxies' molecular gas within the interstellar medium (ISM) and their current star formation rate (SFR). Although the process of star formation is complicated, empirical relationships between galaxies' stellar masses, gas masses, metallicities, SFRs, etc. have been well studied, particularly in the local Universe. 
For instance, originally introduced in terms of volume density by \citet{Schmidt:1959a} and reframed in terms of surface density by \citet{Kennicutt:1998a, Kennicutt:1998b}, the Schmidt-Kennicutt relation delineates a power-law relation  in the $\Sigma_{\rm SFR}$--$\Sigma_{\rm gas}$ plane. 
Measurements of galaxies' SFRs averaged over cosmological scales show a clear peak in star formation activity at $z\sim2$--$3$ with a steep decline to the present day \citep[see e.\/g.\/,][]{Lilly:1996a,Madau:1996a,Hopkins:2004a, Hopkins:2006a, Hopkins:2008a, Madau:2014a}. Therefore, it is critical to study {galaxies' SFRs at the} cosmic peak in order to understand galaxy evolution.

To tie star formation processes to the localised properties of the ISM, it is important to conduct {\it spatially resolved} studies. For galaxies in the local Universe, spatially resolved analyses have been widely implemented for investigating galaxies' gas content and SFRs \citep[e.\/g.\/,][]{Kennicutt:2007a, Bigiel:2008a, Bigiel:2011a, Wei:2010a, Leroy:2013a, Sun:2023a}. However, {while} similar studies are essential for understanding the peak and subsequent decline of cosmic star formation, such work is infrequent due to its difficulty. {Given} their distance, high-redshift galaxies' canonical molecular gas tracer, carbon monoxide (CO), is quite dim (particularly in lower-$J$ transitions) and therefore requires significant integration time to resolve the emission with adequate signal-to-noise ratios (SNRs). {Consequently}, most molecular gas work at high redshift has focused on galaxies' integrated properties \citep[e.\/g.\/,][]{Buat:1989a, Kennicutt:1989a, Kennicutt:1998a, Daddi:2010a, Genzel:2010a, Tacconi:2013a}.

Despite these challenges, there have been a handful of studies focusing on the gas properties of high-redshift galaxies {at (approximately) kiloparsec and sub-kiloparsec resolution} \citep[e.\/g.\/,][]{Sharon:2013a, Genzel:2013a, Rawle:2014a, Hodge:2015a, Canameras:2017a, Tadaki:2018a, Gomez:2018a, Sharon:2019a, Cochrane:2021a, Ikeda:2022a, Nagy:2022a, Tsukui:2023a}. Unfortunately, the resulting measurements of the Schmidt-Kennciutt relation are quite mixed, likely due to the heterogeneous nature of (1) the galaxy types observed (including ``Main Sequence", dusty starburst, and AGN-host galaxies), (2) the observational tracers used to measure gas masses and SFRs (CO $J_{upper}=1$--$7$ lines, and either obscured or unobscured SFR tracers), and (3) the redshifts of the sources being studied ($z\sim1.5$--$5$). The range of Schmidt-Kennicutt indices (correlation slopes) is $n=1.0$--$2.0$, which is biased quite high relative to the clustering around $n=1.0$ for local galaxies \citep{Leroy:2013a}. Many of the high-redshift systems {studied so far are found to have} higher star formation efficiencies (SFEs{, the SFR per unit gas mass) than galaxies on} the (redshift-appropriate) Main Sequence, but it is hard to disentangle how much of the offset is due to the starburst nature of the observed galaxies vs. the different choice (though often well-justified) in CO-to-H$_2$ conversion factor \citep[e.\/g.\/,][]{Daddi:2010a,Genzel:2010a}. {Even in the case where the galaxies have been well-resolved on (sub-)kiloparsec scales, understanding} how these different observations and choices affect the inferred Schmidt-Kennicutt relation at high redshift is clearly necessary.

In order to constrain the effects of observational tracer choice in measuring the Schmidt-Kennicutt relation at high redshift, we have observed a single galaxy using three SFR tracers (two of which are new observations), and two new molecular gas tracers (two others had been published earlier). The galaxy we observed is SDSS J090122-37+181432.3 (hereafter: J0901), a $z=2.26$ UV-bright galaxy lensed by a group of galaxies that has a luminous red galaxy at $z=0.35$, which was first investigated by \citet{Diehl:2009a}, \citet{Hainline:2009a}, and \citet{Fadely:2010a}. J0901 is lensed into three components: a northern and southern arc (the northern arc being a merged pair of partial images of J0901), and a fainter counterimage in the west. J0901 contains a weak AGN, as indicated by the [N\,II]/H$\alpha$ ratios in the two bright arcs, with evidence of outflows given the large line widths seen in \citet{Genzel:2014a}. \citet{Saintonge:2013a} conducted {IR spectral energy distribution (SED)} fitting across {\it Herschel}/PACS and SPIRE bands, deriving a total IR {(TIR, $8$--$1000\,{\rm \mu m}$)} luminosity $L_{\rm TIR}=1.80^{+0.42}_{-0.41}\times 10^{12}L_\odot$ and an implied $\rm SFR_{TIR}=268^{+63}_{-61}M_\odot\,yr^{-1}$ \citep[both corrected for the updated magnification factor of $\mu=30$ from][hereafter: S19] {Sharon:2019a}. These works \citep[and others e.\/g.\/,][]{Rhoads:2014a,Liu:2023a} find that J0901 is a massive rotating disk galaxy that is on the Main Sequence. Prior measurements of the Schmidt-Kennicutt relation in \citetalias{Sharon:2019a} shows that J0901 is a normal star-forming galaxy, with $M_{\rm gas}=(1.6^{+0.3}_{-0.2})\times10^{11}\,{M_\sun}$, and $M_\star=(9.5^{+3.8}_{-2.8})\times10^{10}\,{M_\sun}$ \citep[for an assumed Kroupa IMF,][]{Kroupa:2001a}. The authors find a super-linear index of $n\sim1.5$--$1.8$ (depending on how much emission is resolved out) with no significant dependence on whether the CO(1--0) or CO(3--2) line is used as the gas tracer. However, the ${\rm H\alpha}$ line was used to trace the SFR, and {spatially resolved extinction corrections are important for accurately measuring the SFR surface density and thus the index of the Schmidt-Kennicutt relation in high-redshift galaxies \citep{Genzel:2013a};} given the large dust mass of J0901, correcting for patchy dust obscuration is {likely essential}.

In this work, we present resolved long-wavelength continuum observations at multiple frequencies for the galaxy J0901. We conducted observations using the  VLA at S-band (3\,GHz{, rest-frame 9.8\,GHz}) and L-band (1.4\,GHz{, rest-frame 4.6\,GHz}), and ALMA at {band} 6 (1.2 mm{, rest-frame} 368\,$\mu\rm m${)} of J0901. The VLA observations provide continuum emission measurements at two frequencies, which in addition to the Ka-band observations from \citetalias{Sharon:2019a} further constrain the radio {spectrum} properties of J0901. Moreover, the long wavelength radio emission is a SFR tracer that should be free of dust attenuation effects, and therefore should more accurately reveal the Schmidt-Kennciutt relation than the prior ${\rm H\alpha}$ observations that lacked resolved extinction corrections. The 1.2\,mm continuum map traces the dust emission, and thus provides a complementary map of that obscured star formation. In addition, the ALMA observations measured the CO(7--6) and \ci lines, which provide alternative probes of the molecular gas at different critical densities.

This paper is organized as follows. In Section~\ref{sec:vla_obs} and ~\ref{sec:alma_obs} we introduce the observations and corresponding data reduction, for the VLA and ALMA respectively. In Section~\ref{sec:result} we show the resulting continuum maps and line detections. We then investigate the radio continuum {spectrum} and decompose the thermal and non-thermal emission (Section~\ref{subsec:radio_decomposition}), discuss J0901 in the context of the  {IRRC} (Section~{\ref{subsec:IRRC}}), and check for consistency between different SFR measurements (Section~\ref{subsec:sfr}). We then calculate the gas masses using \ci line and compare them to the value inferred from the 1.2\,mm continuum measurement (Section~\ref{subsec:gasmass}), as well as the inferred molecular gas properties using large velocity gradient (LVG) fits to the CO spectral line energy distribution (SLED; Section~\ref{subsec:SLED}). Finally, in Section~\ref{subsec:skrelation}, we compare the effects of the choice of SFR and gas tracer on the observed Schmidt-Kennicutt relation for J0901. Our results are summarized in Section~\ref{sec:concl}. Throughout this work we adopt $\rm \Omega_\Lambda=0.725$ and $\rm H_0=70.2\,km\,s^{-1}$ (the WMAP7+BAO+$\rm H_0$ cosmology parameters in the $\Lambda$CDM framework; \citealt{Komatsu:2011a}) and the lensing models presented in \citetalias{Sharon:2019a}.

\floattable
\begin{deluxetable}{lcccccl}
\tablewidth{0pt}
\tablecaption{J0901 VLA continuum observations 
\label{tab:j0901obs}}
\tablehead{ {Band} & {Configuration} & {Date} & {$N_{\rm Ant}$} & {$t_{\rm track}$} & {$t_{\rm target}$} \\
{$\nu_{\rm obs}$} & {} & {} & {} &{(hour)} &{(hour)} 
}
\startdata
{S-band} & {B} & 2020 June 24 & 27 & 1.3 & 0.8 \\ 
{${3\,{\rm GHz}}$} & {} & 2020 June 26 & 27 & 2 & 1.5 \\ 
{} & {} & 2020 June 27 & 27 & 2 & 1.4 \\ 
{} & {} & 2020 June 28 & 26 & 2 & 1.5 \\ 
\hline
{S-band} & {A} & 2022 May 10 & 26 & 2 & 1.4 \\ 
{${3\,{\rm GHz}}$} & {} & {} &{}&{}\\
\hline
{L-band} & {A} & 2022 May 28 & 26 & 2 & 1.5 \\ 
{${1.4\,{\rm GHz}}$} & {} & 2022 June 19 & 26 & 1 & 0.5 \\
{} & {} & 2022 June 20 & 25 & 1.5 & 1 \\ 
{} & {} & 2022 June 25 & 27 & 2 & 1.5 \\ 
{} & {} & 2022 June 28 & 27 & 2 & 1.5 \\ 
{} & {} & 2022 July 01 & 27 & 2 & 1.5 \\ 
{} & {} & 2022 July 01 & 27 & 2 & 1.5 \\ 
{} & {} & 2022 July 03 & 27 & 1.5 & 1 \\ 
{} & {} & 2022 July 03 & 27 & 2 & 1.5 \\ 
\enddata
\end{deluxetable}

\section{VLA Continuum Observations \& data reduction}\label{sec:vla_obs}

\begin{figure*} 
	\centering
	\vspace{0cm}
	\subfigtopskip=0pt 
	\subfigbottomskip=0pt 
	\subfigcapskip=0pt
    \subfigure{
		\label{fig:Lband_rms}
		\includegraphics[width=0.49\linewidth]{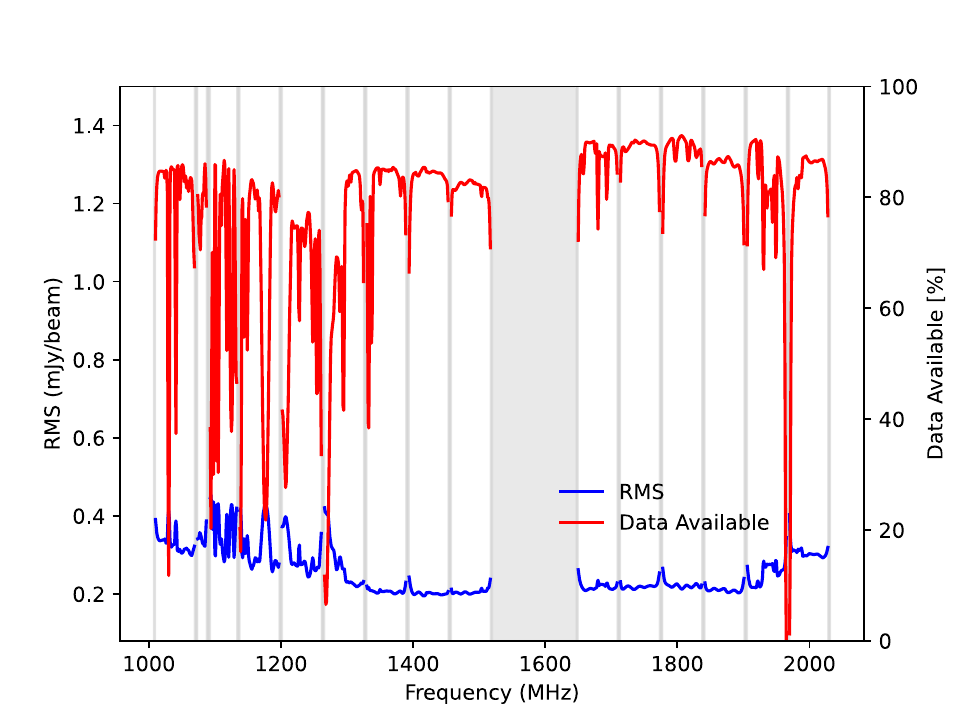}}
	\subfigure{
		\label{fig:Sband_rms}
		\includegraphics[width=0.49\linewidth]{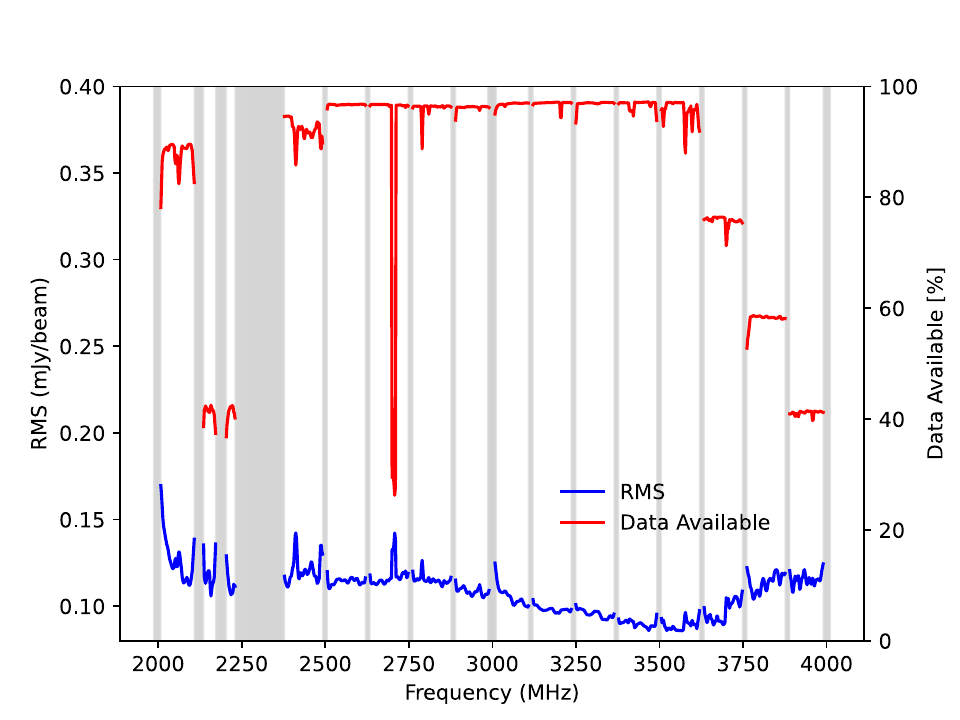}}
	\caption{The channel-based RMS (blue, left axes) and available data percentage (red, right axes) after reduction of the L-band  {(${1.4\,{\rm GHz}}$)} data (left) and S-band  {(${3\,{\rm GHz}}$)} data (right). Gray-shaded regions highlight channels that have been completely flagged, which generally happens at spectral window edges where the noise is considerably higher.}
	\label{fig:flag_ratio}
\end{figure*}

Our VLA observations were taken in S-band ($\nu_{\rm obs}=3\,{\rm GHz}$) and L-band ($\nu_{\rm obs}=1.4\,{\rm GHz}$). We first obtained 7.3 hrs of ${3\,{\rm GHz}}$ data in 2020 (ID: 20A-289; PI C. Sharon), carried out in the B-configuration. We obtained a second round of observations in 2022 (ID: 22A-157; PI C. Sharon) which included 2 hours  {at ${3\,{\rm GHz}}$} and 16 hours  {at ${1.4\,{\rm GHz}}$}, both in the A configuration. We summarise the VLA continuum observations in Table~\ref{tab:j0901obs}. The frequency based RMS and percentage of unflagged data after reduction is shown in Fig.~\ref{fig:flag_ratio}.

\subsection{VLA S-band {(${\nu_{\rm obs}=3\,{\rm GHz}}$)} Observations}\label{subsec:vla_obs_s}

With the standard 8-bit sampler set, we chose two adjacent 1\,GHz basebands centred at 2.5 and 3.5\,GHz. This setup makes for continuous frequency coverage of 2\,GHz centred at 3\,GHz (rest-frame 9.8\,GHz). As we do not require polarization information, the WIDAR correlator was set in the ``OSRO Dual Polarization" mode. The 2\,GHz bandwidth is split into 16 spectral windows with 64 channels of 2\,MHz each. The pointing centre is $\alpha = 09^h 01^m 22^s .45, \delta = +18\degree 14\arcmin 31\arcsec .5$, roughly in the middle of J0901's three arcs. The observations were scheduled in one 1.3 hr track and four 2 hr tracks (total on-source integration time is 6.6 hr). One track of the longer two (1.4 hrs on-source) was taken in the A configuration, and the others in the B configuration. We used 3C\,147 for bandpass and flux {scale} calibrations, and J0854+2006 to calibrate time variations in amplitude and phase. At the beginning of each track, we observed 3C\,147, then alternated between J0854+2006 ($\sim$6 min) and J0901 ($\sim$13 min) until the end of the track. {For flux density measurements, we assume the standard 5\% calibration uncertainty in this band as given by the VLA Observational Status Summary.
}

The data reduction involved running the standard pipeline with additional manual intervention for necessary data quality flagging. We used the CASA \citep{McMullin:2007a} pipeline appropriate for the given observation date (for the A-configuration: CASA version 6.2.1.7, Pipeline version 2021.2.0.128; for the B-configuration: CASA version 5.4.1-32, Pipeline version 42270). While the data reduction was successful according to the pipeline quality assessment (QA) reports, there was still considerable RFI present in the data. We further clipped those with amplitudes higher than 0.3\,Jy, and then conducted a series of very careful flagging runs based on amplitude vs. $uv$-plane distance and amplitude vs. channel in each spectral window. In the end, less than half of the original data for 2116--2244\,MHz and 3884--4012\,MHz remained, while the 2244--2372\,MHz range was totally flagged. These spectral windows are seriously affected by satellite downlink communications and digital audio radio satellites \footnote[1]{https://science.nrao.edu/facilities/vla/observing/RFI/S-Band}. Most of the other frequencies are only mildly flagged.

The S-band {(${\nu_{\rm obs}=3\,{\rm GHz}}$)} fractional bandwidth (bandwidth divided by the band centre frequency) is 0.67, which requires wide band imaging procedures for continuum image reconstruction. We use the multi-term multi-frequency synthesis algorithm to account for spatial variations of spectral indices \citep{Rau:2011a}, which is implemented using \texttt{mtmfs} as the deconvolver and setting Taylor term as 2 in the CASA {\sc tclean} task. We use the H{\"o}gbom algorithm \citep{Hogbom:1974a} to derive clean models and deconvolve down to $2\,\sigma$.
The imaging is implemented on $2048\times 2048$ pixels, each with a size of $0.\arcsec2\times 0.\arcsec2$.  {Since J0901 is $\sim$20}\arcsec {across, it only occupies} the central $\sim$ {0.2\% of the area we image}. The large imaging field of view is necessary to ensure that all emission from bright nearby sources' sidelobes are removed during cleaning and thus that we obtain the correct flux {density} for J0901.

We experimented with various $uv$-weighting schemes in order to determine the best compromise between sensitivity and resolution. We investigated natural, uniform, and Briggs \citep{Briggs:1995a} weighting with \textit{robust} equal to 0.8, 0.5, and 0 (hereafter expressed as `Briggs'+robust) respectively. Briggs0 and uniform weighting result in maps whose source region SNR is low, likely due to the limited sensitivity of the single track in A configuration, and were therefore discarded from further analysis. Briggs0.8 weighting leads to a very similar map compared to that for natural weighting. We hereby focus on the natural and Briggs0.5 weighting in the subsequent analyses.

Emission from several nearby sources needed to be accounted for during imaging. The very bright radio source 3C\,213  {\citep[0.62\,Jy at ${3\,{\rm GHz}}$,][]{Wright:1990a}} is outside the field of view (though see Section~\ref{subsec:vla_obs_l}, where its effects are more significant), and turned out to only have a mild effect on the  {${3\,{\rm GHz}}$} image quality. However, the radio galaxy [CCH85] 0857+1844A (hereafter: CCH85) at $\alpha=09^h 00^m 48^s .12$, $\delta= +18\degree 32' 40'' .7$, is only $\sim 2.\arcmin6$ away from our target. It mildly contaminates our target in the Briggs0.5 map, but almost totally overwhelms J0901 when we use natural weighting. It has no {${3\,{\rm GHz}}$ flux density measurements in the} literature, but the 1.4\,GHz flux {density} is $\rm 333.0\pm16.3\ mJy$ \citep{Coleman:1985a}---about three orders of magnitude higher than that of J0901. To correct for the influence of CCH85, we implement a one-cycle amplitude+phase self-calibration. For Briggs0.5 weighting, the sidelobes are successfully removed after self-calibration, and the resulting map has a uniform noise distribution. For the natural weighting images, the self-calibration leaves small (below $2\,\sigma$) but noticeable ripples from the sidelobes in the southwestern part of J0901. Following \citet{Rau:2011a}, we used the \texttt{multiscale} algorithm (with scales of ``0" for a point-like component, $\rm 2\times beam$ and $\rm 5\times beam$) and a higher Taylor term of 4, since the map dynamic range is high and the source is resolved with complex spectral indices. While these changes do not make noticeable differences to the Briggs0.5 map, for the natural weighting map, the remaining ripples near J0901 are removed. Lastly, we correct the primary beam variation with the CASA task \texttt{widebandpbcor}, though J0901 only occupies a small portion near the centre of the field of view and the correction makes little difference (the primary beam response is 0.9996 at the distant edge of J0901 arcs).

The surface brightness of the central galaxies lensing J0901 are typically $\sim 1\,{\rm mJy\,beam^{-1}}$---more than 30 times the peak brightness of J0901 itself. In order to check whether J0901 is free of contamination from the foreground lens, we also made images with the lens's emission removed using $uv$-plane continuum subtraction in CASA. We construct the $uv$-plane model by cleaning within a manually defined mask that encloses only the central lensing galaxies (based on the map generated in the previous step). We then run \texttt{uvsub} to subtract the clean algorithm's model of the lens from the data, followed by a normal deconvolution process using the parameters described above without continuum subtraction (clean or \texttt{multiscale}, depending on the weighting). For the Briggs0.5 weighting, $uv$-plane subtraction of a model derived from cleaning to a $2\,\sigma$ threshold is very successful, with few noticeable residuals. For natural weighting, we instead clean the central lens down to the $1\,\sigma$ level. Even with this deeper subtraction, the residual emission from the lens is comparable to the surface brightness of the western image in J0901.

Although natural weighting should produce the best SNR maps, ultimately the Briggs0.5 map has lower noise ($\rm 3.0\,\mu Jy\, beam^{-1}$) than the map with natural weighting ($\rm 3.5\,\mu Jy\, beam^{-1}$). The higher noise in the latter is likely due to residual sidelobes of other nearby bright sources, although these do not stand out during visual inspection. The Briggs0.5 map has a resolution of $1.\arcsec2\times 1.\arcsec1$, which is considerably better than that of the natural weighting map ($1.\arcsec8\times 1.\arcsec7$). We therefore use the Briggs0.5 maps for the subsequent analyses unless specified otherwise in the text. As discussed in Section~\ref{sec:result}, there is little difference between the integrated  {flux densities} of J0901's images for the different weighting choices.

\subsection{VLA L-band {(${\nu_{\rm obs}=1.4\,{\rm GHz}}$)} Observations}\label{subsec:vla_obs_l}

Our L-band {(${\nu_{\rm obs}=1.4\,{\rm GHz}}$)} observations include two 512\,MHz basebands that provide continuous frequency coverage from 1\,GHz to 2\,GHz. The instrumental set-up is generally the same as that of the  {${3\,{\rm GHz}}$} observations, with the standard 8-bit sampler, the ``OSRO Dual Polarization" mode for the WIDAR correlator, and 16 spectral windows of 64$\times$ 1\,MHz channels. We obtained nine tracks (six 2-hour blocks, two 1.5-hour blocks, and one 1-hour block) in A configuration for a total of 11.5 hours on-source. We use the same pointing centre as for the  {${3\,{\rm GHz}}$ observations}. We again use 3C\,147 and J0854+2006 as the flux {scale}/bandpass and amplitude/phase calibrators, respectively, alternating between J0854+2006 and J0901 in the same way as for the  {${3\,{\rm GHz}}$} observations. For flux {density} measurements, we  {assume the standard 5\% calibration uncertainty in this band as given by the VLA Observational Status Summary}.

We first implemented the standard data reduction using CASA version 6.2.1.7 and pipeline version 2021.2.0.128. The QA reports show that the data reduction was successful. We then perform manual flagging to remove additional RFI. We totally remove the two spectral windows covering 1520--1648\,MHz due to their high noise; this frequency range is used by multiple satellite constellations (GPS, GLONASS, INMARSAT, etc.\footnote[2]{https://science.nrao.edu/facilities/vla/observing/RFI/L-Band}) and high-altitude balloons. We summarise the channel-based flagging fractions and resulting RMS values for the remaining  {${1.4\,{\rm GHz}}$} visibilities in left panel of Fig.~\ref{fig:flag_ratio}.

For imaging, the multi-term modeling algorithm is again necessary due to the large fractional bandwidth (1/1.5). We conduct H{\"o}gbom cleaning using the \texttt{mtmfs} as deconvolver with a Taylor term of 2 in CASA's {\sc tclean}. We tested weighting schemes including natural, Briggs0.8, Briggs0.5, Briggs0, and uniform. For all weighting schemes we set the cleaning threshold to $2\,\sigma$. To ensure adequate sampling of the synthesized beam, we use $0.\arcsec2\times 0.\arcsec2$ as the pixel size.

As with the ${3\,{\rm GHz}}$ data, the ${1.4\,{\rm GHz}}$ Briggs0.8 maps are very similar to the maps with natural weighting. The Briggs0 and uniform weightings result in very low SNRs, so we again only focus on the natural and Briggs0.5 weightings. A good approximation between ${1.4\,{\rm GHz}}$ and Q-band {(${\nu_{\rm obs}=45\,{\rm GHz}}$)} for the primary beam full-width half maximum\footnote[3]{https://science.nrao.edu/facilities/vla/docs/manuals/oss/performance/fov} (in arcmin) is $\theta_{\rm PB}=42/\nu_{\rm GHz}$. The field of view  {at ${1.4\,{\rm GHz}}$} is therefore four times larger than that  {at ${3\,{\rm GHz}}$}, leaving the imaging more easily affected by sources at a large angular separation. Interestingly, the radio galaxy CCH85 does not significantly affect J0901  {at ${1.4\,{\rm GHz}}$} for either natural or Briggs0.5 weighting. However, the larger field of view encloses the strong radio source 3C\,213 at $\alpha=09^h 00^m 48^s .373$, $\delta= +18\degree 32' 12'' .8$. The flux {density} of this source is $\rm 1.23\pm0.04\ Jy$ according to the NVSS catalogue \citep{Condon:1998a}, with similar measurements seen in \citet{Wright:1990a}, \citet{White:1992a}, and \citet{Pauliny-Toth:1966a}. Its flux density is about five orders of magnitude larger than the peak flux density of J0901. Although it is $\sim 19\arcmin$ away from our source, where the primary beam response is about only $\sim 0.23$, 3C\,213 severely affects the central region of the image when we use natural weighting. We therefore image an extremely large area of $16384\times 16384$ pixels to capture its effect. 
In contrast to the ${3\,{\rm GHz}}$ data, we were unable to self-calibrate the influences from 3C\,213 without introducing additional artifacts (using either weighting scheme). For the Briggs0.5 weighting, however, 3C\,213 sidelobes are quite weak (below $1\,\sigma$) in the source region even though no self-calibration is implemented. 
In consequence, we only keep the Briggs0.5 weighting result without self-calibration or multiscale cleaning, which yields a spatial resolution of $1.\arcsec1\times 1.\arcsec0$ and RMS level of $\rm 4.0\,\mu Jy\, beam^{-1}$. We then apply a primary beam correction which ultimately makes almost no change in the small central fraction of the image occupied by J0901.

Following the same strategy used for the ${3\,{\rm GHz}}$ data, we then subtract a visibility-based model of the central lens based on a round of $2\,\sigma$ cleaning that excludes J0901 itself. This $uv$-plane continuum subtraction successfully removes the lens emission without significantly affecting regions closer to J0901 itself.

\section{ALMA Observations \& Data Reduction}\label{sec:alma_obs}

J0901 was observed during ALMA Cycle 2 with both the 12m Array and Atacama Compact Array (including both the 7m and Total Power Arrays) as part of Project \#2013.1.00952.S (PI: Sharon). Observations using the 12m Array were taken on 2014 April 30 for a total of $34\,{\rm minutes}$ on source. Baseline lengths for the then-available 36 antennas ranged from $15.1$--$348.5\,{\rm m}$. Observations using the 7m Array were taken on 2014 April 30, May 2, and May 19 for a total of $66\,{\rm minutes}$ on source. Total Power observations were taken nine times between 2014 Dec 27 and 2015 Jan 15 for a total of $5.5\,{\rm hours}$ on source. The correlator was configured using overlapping spectral windows in each sideband. We centered the CO(7--6) line at $247.5\,{\rm GHz}$ (observed frequency {, or ${\lambda_{\rm obs}\sim1.2\,{\rm mm}}$; band} 6) in the lower spectral window of the upper sideband, allowing for coverage of the \ci line at higher frequencies ($248.3\,{\rm GHz}$, in the overlap region of two spectral windows). The lower sideband was centered at $232.9\,{\rm GHz}$ to capture continuum emission. Due to the expected width of the lines, we used the largest available channel size in Frequency Domain Mode with only two polarizations, which is $488.281\,{\rm kHz}$ ($\sim0.6\,{\rm km\,s^{-1}}$), for the upper-sideband; for the continuum observations in the lower-sideband, we used the Time Domain Mode channel size of $15.625\,{\rm MHz}$. We used observatory-selected calibrators and observation sequences. For the 12m Array, Ganymede was used for amplitude and flux calibration, J0750+1231 for bandpass and pointing calibration, and J0854+2006 for phase calibration for the 12m Array. We assume {the standard} 10\%   flux scale uncertainties for  {band 6 (${1.2\,{\rm mm}}$), as given by the annual call for proposals}. For the 7m Array observations, either Pallas or Ceres was used for flux calibration, J0854+2006 was used for phase calibration, and either J0854+2006 or J1058+0133 was used for bandpass calibration. For the Total Power observations, J0854+2006 was used for all calibrations. 

For calibration of the 12m Array data, we used a  {script provided by the North American ALMA Science Center} and CASA version 4.5.0 in order to more closely examine the outputs of the pipeline calibration. This examination resulted in shortening the amplitude/phase solution interval relative to the pipeline default and the removal of data from two antennas in order to improve calibration (in addition to the flagging of a single channel affected by atmospheric absorption identified during the default pipeline quality analysis). For calibration of the 7m Array data, we used the observatory-generated calibration script for CASA version 4.2.2, since additional inspection resulted in no calibration improvement. For the Total Power data, we used the observatory-provided manually calibrated data produced in CASA version 4.5.0. As neither of the spectral lines were detected, and the Total Power  {array is not sensitive to} continuum emission, we did not  {include} the Total Power data  {when making} our final maps. We used natural weighting to produce the final maps (in CASA 4.5.0), which resulted in synthesized beam sizes of $1.\arcsec6\times1.\arcsec0$ at a position angle of $58^{\degree}$ for the two emission lines, and $1.\arcsec7\times1.\arcsec1$ at a position angle of $58^{\degree}$ for the continuum map. We binned the frequency axis into $14\,{\rm km\,s^{-1}}$ ($11.5\,{\rm MHz}$) channels, and our final average per-channel noise is $0.47\,{\rm mJy\,beam^{-1}}$ for the CO(7--6) line and $0.61\,{\rm mJy\,beam^{-1}}$ for the \ci line after $uv$-plane continuum subtraction in CASA (continuum subtraction was performed separately for the 12m and 7m Array data). The ${1.2\,{\rm mm}}$ continuum map uses only the lower-sideband spectral windows (since they provided more than adequate S/N and are widely separated from the upper-sideband) for a final map noise of $35\,{\rm \mu Jy\,beam^{-1}}$ over the nearly $4\,{\rm GHz}$ of bandwidth.

\section{Results}\label{sec:result}

\subsection{Continuum detections}\label{subsec:cont}

We successfully detected all three arcs of J0901 in the VLA 1.4\,GHz and 3\,GHz continuum maps, as well as with ALMA at 1.2\,mm. 
We list the flux {density} measurements and map properties in Table~\ref{tab:maps_flux}, for our fiducial Briggs0.5 weighting maps, as well as for the naturally-weighted maps, both with and without $uv$-continuum subtraction of the central lens. The reported error bars include both the statistical uncertainties and  {the appropriate flux scale} calibration uncertainty.

\floattable
\begin{deluxetable*}{ccccccccc}
\rotate
\tablewidth{0pt}
\tablecaption{J0901 continuum flux density measurements
\label{tab:maps_flux}}
\tablehead{ {Band} & {$\nu_{\rm obs}$, $\nu_{\rm rest}$} & {Map} & {$\theta_M\times\theta_m$} & {RMS} & {$S_{\rm north}$} & {$S_{\rm south}$} & {$S_{\rm west}$} & {$S_{\rm total}$}\\
{}  & (GHz, GHz) & {}  & ($\arcsec$) & {$(\rm\mu Jy\ beam^{-1})$} & {$(\rm\mu Jy)$} &{$(\rm\mu Jy)$} &{$(\rm\mu Jy)$} & {$(\rm\mu Jy)$}
} 
\startdata
S-band & {3, 9.8}& Natural & $1.8\times 1.7$ & 3.5 &  ${130\pm16}$ &  ${126\pm13}$ &  ${64\pm9}$ & ${320\pm26}$\\
S-band & {3, 9.8}& Natural-uvsub & $1.8\times 1.7$ & 3.5 & ${184\pm18}$ &  ${118\pm13}$ &  ${66\pm9}$ & ${367\pm28}$\\
S-band & {3, 9.8}& Briggs0.5 & $1.2\times 1.1$ & 3.0 &  ${168\pm19}$ &  ${120\pm14}$ &  ${47\pm9}$ &  ${334\pm28}$\\
S-band & {3, 9.8}& Briggs0.5-uvsub & $1.2\times 1.1$ & 3.0 &  ${162\pm18}$ &  ${120\pm14}$ & ${46\pm9}$ &  ${328\pm28}$\\
\hline
L-band & {1.4, 4.6}& Briggs0.5 & $1.1\times 1.0$ & 4.0 &  ${222\pm27}$ &  ${170\pm21}$ &  ${112\pm14}$ &  ${505\pm42}$\\
L-band & {1.4, 4.6}& Briggs0.5-uvsub & $1.1\times 1.0$ & 4.0 &  ${226\pm27}$ & ${170\pm21}$ & ${107\pm14}$ & ${503\pm42}$\\
\hline
ALMA  {band} 6 & {250, 814}& Natural & $1.7\times 1.1$ & 30.0 & $(6.69\pm0.69)\times 10 ^3$ & $(5.05\pm0.52)\times 10^3$ & $(2.25\pm0.25)\times 10^3$ & $(14.00\pm1.42)\times 10^3$
\enddata
\end{deluxetable*}

\begin{figure*} 
	\centering
	\vspace{0cm}
	\subfigtopskip=0pt 
	\subfigbottomskip=0pt 
	\subfigcapskip=0pt
	\subfigure{
		\label{fig:Sband-natural}
		\includegraphics[width=0.49\linewidth]{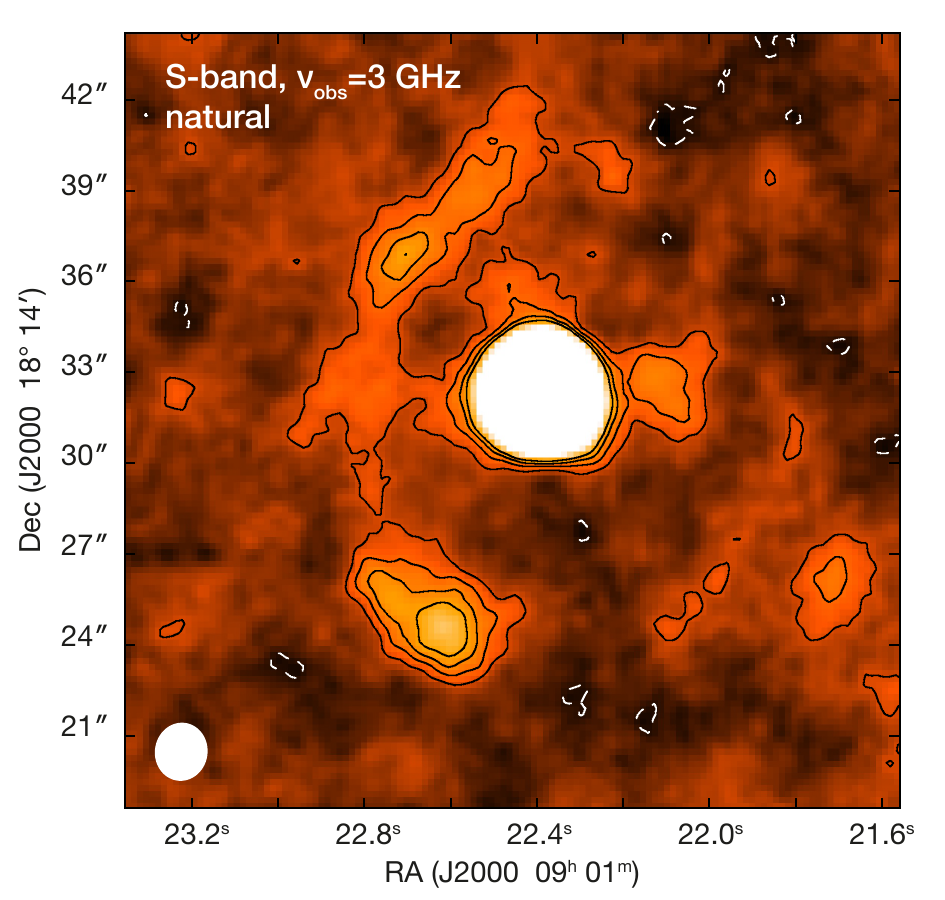}}
	\subfigure{
		\label{fig:Sband-natural-uvsub}
		\includegraphics[width=0.49\linewidth]{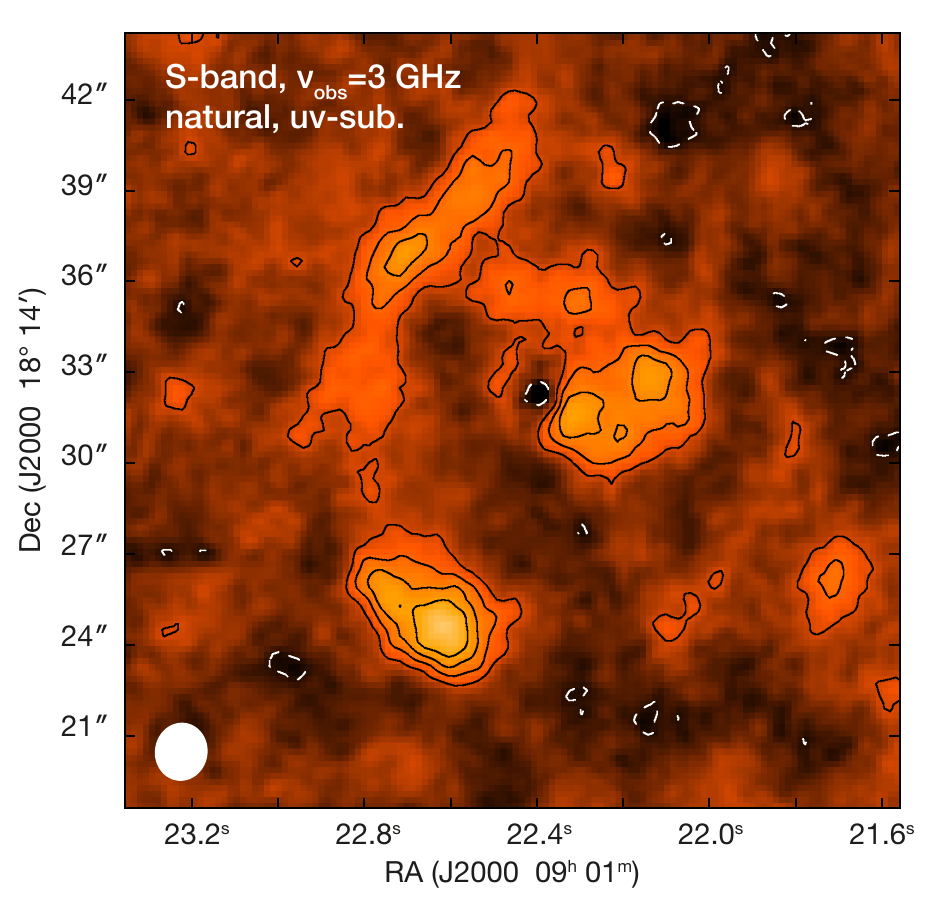}}
	\subfigure{
		\label{fig:Sband-briggs0.5}
		\includegraphics[width=0.49\linewidth]{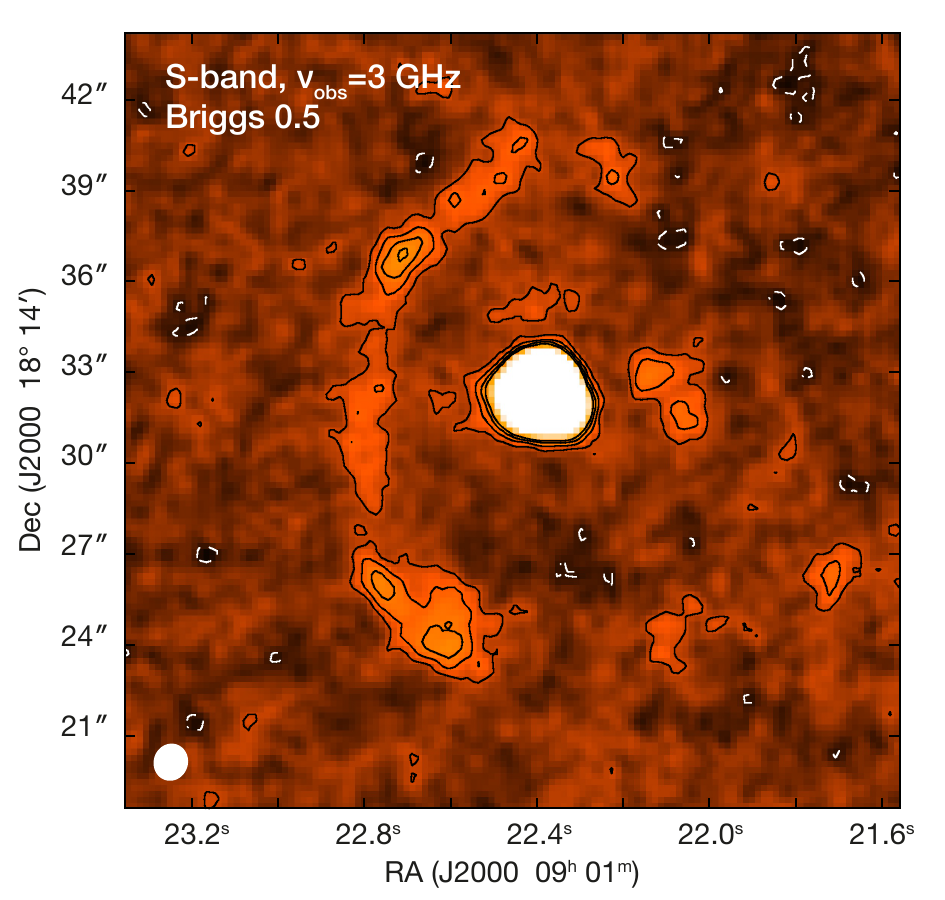}}
	\subfigure{
		\label{fig:Sband-briggs0.5-uvsub}
		\includegraphics[width=0.49\linewidth]{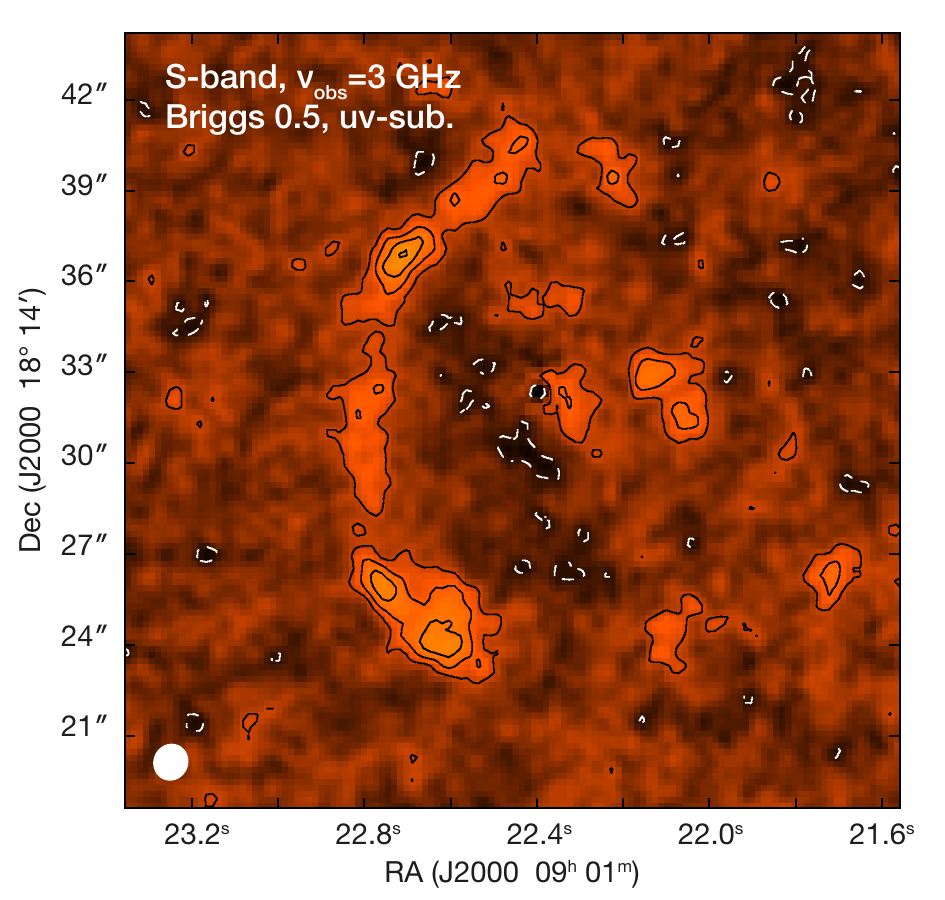}}
	\caption{Continuum maps from VLA S-band  {(${\nu_{\rm obs}=3\,{\rm GHz}}$)} observations, for the naturally weighted map (upper left),  the naturally weighted map after subtracting continuum from the central lens (upper right), the Briggs0.5 weighted map (lower left), and the Briggs0.5 weighted map after subtracting the central lens emission (lower right). Contour levels are $(-2, 2, 4, 6, 8)\times$ the RMS values given in Table~\ref{tab:maps_flux}. Positive and negative contours are distinguished with black and (dashed) white colours, respectively. We set a maximum threshold of 50 $\mu\rm Jy\,beam^{-1}$ to the colour scale so that the bright central lens is ignored and the J0901 flux  {density} distribution is more visible. The synthesized beams are shown in white in the lower left corner of each panel.}
	\label{fig:Sband_flux}
\end{figure*}

We show the resulting {${3\,{\rm GHz}}$} continuum (and lens-removed) maps with natural and Briggs0.5 weighting in Fig.~\ref{fig:Sband_flux}. We resolve the three arcs at $\geq4\,\sigma$ significance. With the better resolution of the Briggs weighting, the southern and western arcs are resolved to have two separated sub-regions at the $6\,\sigma$ and the $4\,\sigma$ significance respectively. We find that the flux {densities} measured from the two weighting schemes are consistent within the uncertainties. {Those} measured from the maps (using apertures large enough to enclose all the emission) with $uv$-plane removal of the central lens are consistent with those without central lens subtraction, suggesting that our method for continuum subtraction was largely successful (albeit with less clear separation for the western image in the natural weighting map). Given this consistency in {flux densities}, we use the Briggs0.5 lens-removed {measurement of ${S_{\rm 3\,GHz}=328\pm28\,{\rm \mu Jy}}$} as our reference measurement.

\begin{figure*} 
	\centering
	\vspace{0cm}
	\subfigtopskip=0pt 
	\subfigbottomskip=0pt 
	\subfigcapskip=0pt
	\subfigure{
		\label{fig:Sflux-conv-briggs0.5}
		\includegraphics[width=0.49\linewidth]{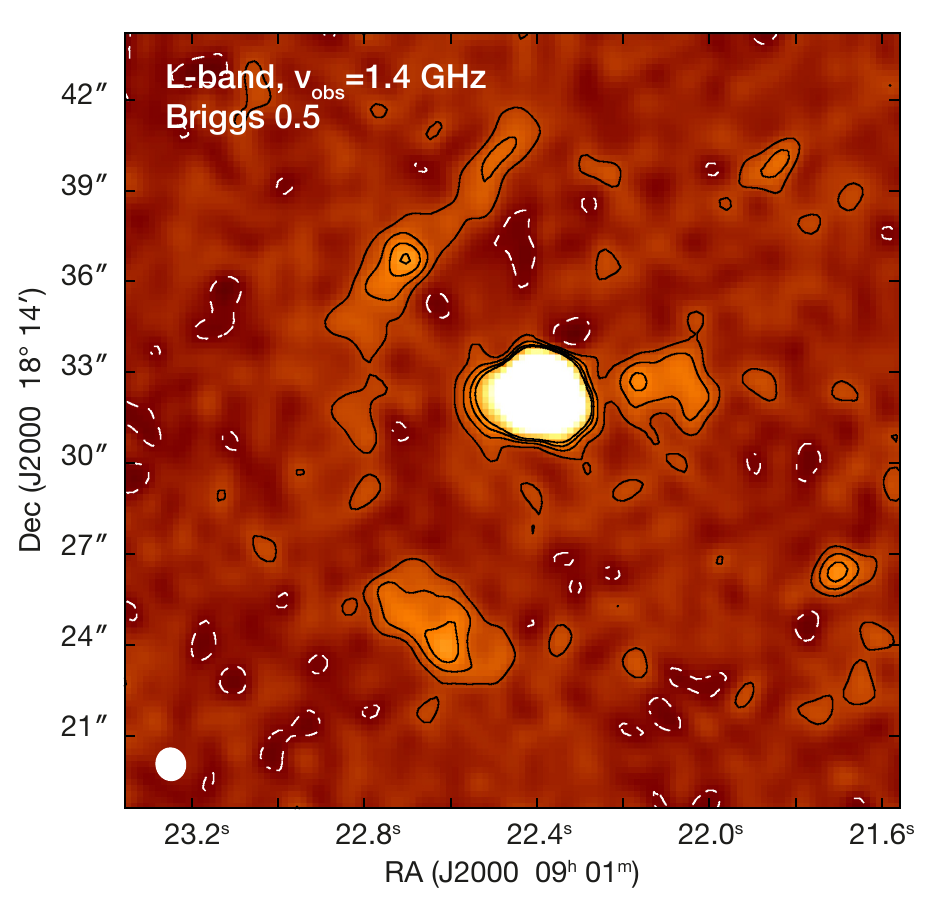}}
	\subfigure{
		\label{fig:Lflux-conv-briggs0.5}
		\includegraphics[width=0.466\linewidth]{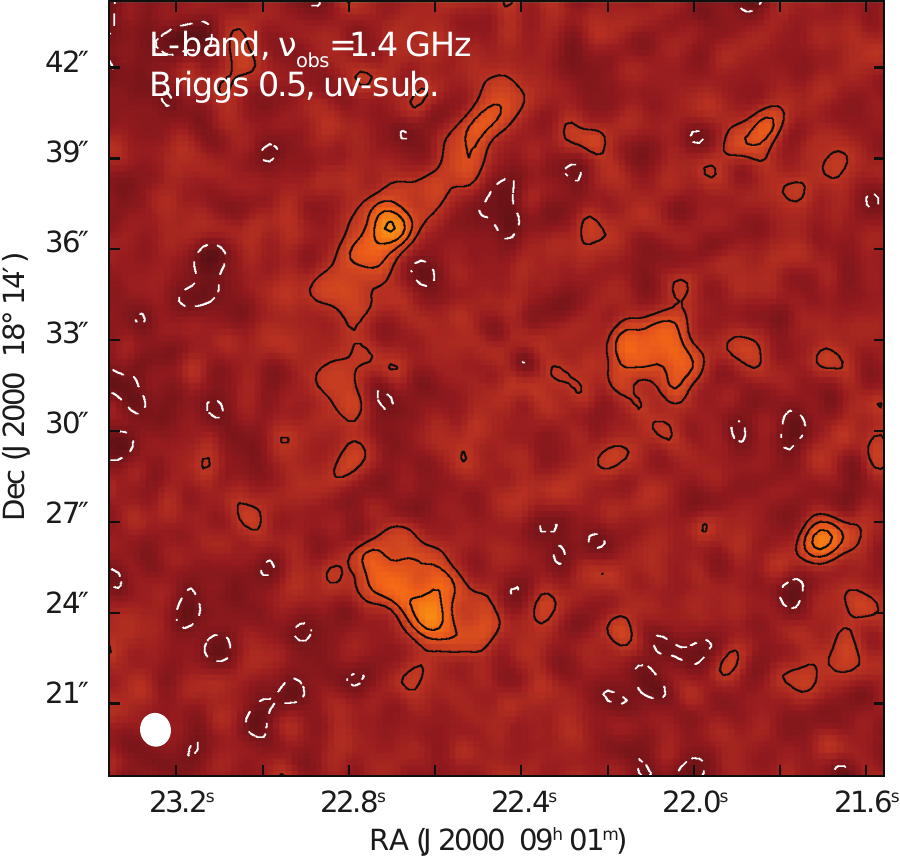}}
	\caption{The VLA L-band {(${\nu_{\rm obs}=1.4\,{\rm GHz}}$)} continuum flux  {density} maps without (left) and with (right) the central lens continuum subtraction. A maximum threshold of 90 $\mu\rm Jy\,beam^{-1}$ is applied to the colour scale so that the bright central lens is ignored and the J0901 flux  {density} distribution is better seen. Contour levels are $(-2,2,4,6,8)\times$ the RMS values given in Table~\ref{tab:maps_flux}. Positive and negative contours are distinguished with black and (dashed) white colours, respectively. The synthesized beams are shown in white at the lower left corners.}
	\label{fig:Lflux-briggs0.5}
\end{figure*}

We show our final continuum maps at {${1.4\,{\rm GHz}}$} in Fig.~\ref{fig:Lflux-briggs0.5}. As discussed in Section~\ref{subsec:vla_obs_l}, we only use the Briggs0.5 flux {density} maps due to the difficulty of removing contamination from nearby sources' bright sidelobes. All three arcs of J0901 are clearly seen, with {peaks} all exceeding $6\,\sigma$. The integrated  {flux densities} measured from the Briggs0.5 maps with the central lens included and removed are in excellent agreement. To be consistent with our {${3\,{\rm GHz}}$} analysis, we use the Briggs0.5 lens-removed flux density of  {${S_{\rm 1.4\,GHz}=503\pm42\,{\rm \mu Jy}}$} as our reference measurement.

\begin{figure}
\centering
    \includegraphics[width=0.95\columnwidth]{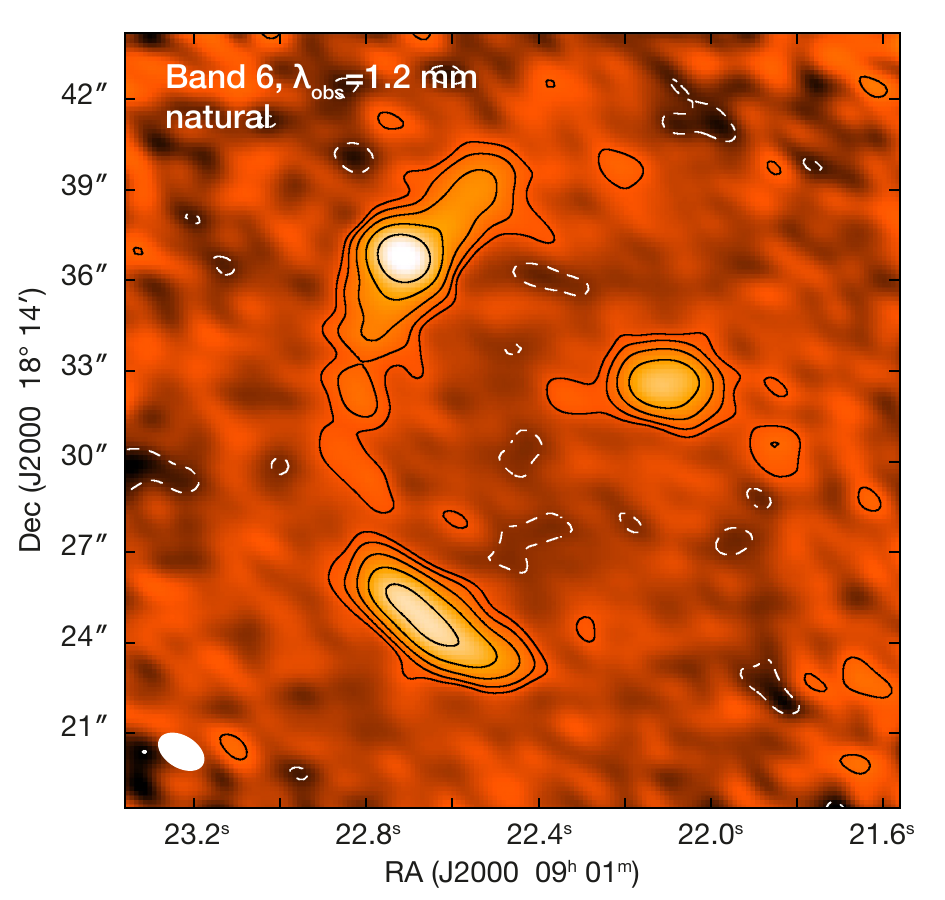}
    \caption{The ALMA {band 6 (${\lambda_{\rm obs}=1.2\,{\rm mm}}$} continuum map produced using natural weighting. Contour levels are set to be $(-2,2,4,8,16,32,64)\times$ the RMS value given in Table~\ref{tab:maps_flux}. Positive and negative contours use black and (dashed) white colors, respectively. The synthesized beam is shown in white at the lower left.}
    \label{fig:alma1.2mm}
\end{figure}

We show the 1.2\,mm ALMA continuum map in Fig.~\ref{fig:alma1.2mm}. The 1.2\,mm (rest $0.37\,{\rm mm}$) wavelength traces the dust emission of the galaxy, which is significantly brighter than the VLA continuum observations, and therefore we clearly detect all three images with peak SNR=78. The central lensing galaxy is not seen at this wavelength and therefore does not require additional $uv$-plane continuum subtraction. The total observed continuum flux {density} of J0901 for the ALMA data is $S_{\rm 1.2\,mm}=14\pm1.4\,{\rm mJy}$, where the uncertainty is dominated by the 10\% flux {scale} calibration uncertainty.

\subsection{Line detections}\label{subsec:lines}

\begin{figure*}
\centering
    \includegraphics[width=2.0\columnwidth]{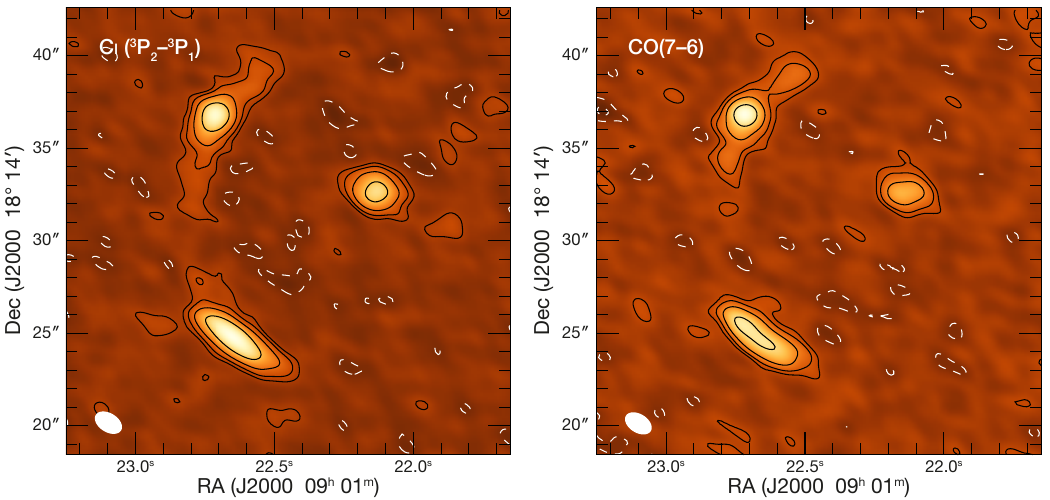}
    \caption{
    Integrated line maps of the \ci (left) and CO(7--6) (right) lines. Both images sum channel maps within the same (rest-frame) $355\,{\rm km\,s^{-1}}$ velocity range. Contours are in powers of $2^n\times\sigma$ (i.\/e.\/, $\pm2\,\sigma$, $\pm4\,\sigma$, $\pm8\,\sigma$) where $\sigma_{\rm CO}=0.04\,{\rm Jy\, km\,s^{-1}\,beam^{-1}}$ and $\sigma_{{\rm C}\textsc{i}}=0.05\,{\rm Jy\, km\,s^{-1}\,beam^{-1}}$. Positive and negative contours use black and white colors, respectively. The synthesized beam FWHMs are shown at lower left.}
    \label{fig:intline}
\end{figure*}

\begin{figure*}
\centering
    \includegraphics[width=2.0\columnwidth]{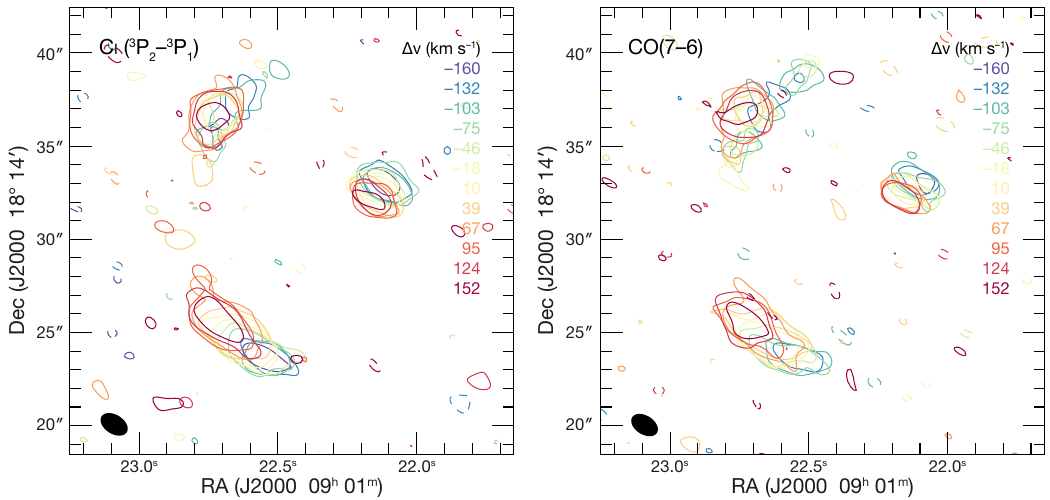}
    \caption{
    Overlaid channel maps of the \ci (left) and CO(7--6) (right) lines. We only show the $\pm3\,\sigma$ contours for clarity (where $\sigma_{\rm CO}=0.7\,{\rm mJy\, beam^{-1}}$ and $\sigma_{{\rm C}\textsc{i}}=0.9\,{\rm mJy\, beam^{-1}}$), which are color-coded by relative velocity; negative contours are dashed. The synthesized beam is shown at lower left.
}
    \label{fig:renzos}
\end{figure*}

\begin{figure*}
\centering
    \includegraphics[width=2.0\columnwidth]{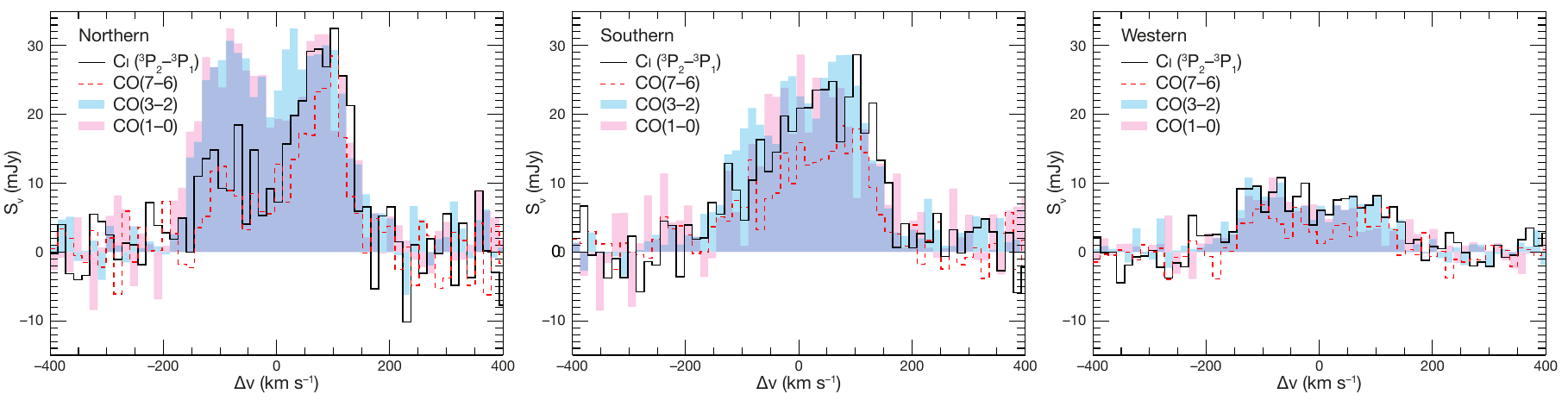}
    \caption{
    ALMA spectra of the \ci (black lines) and CO(7--6) (red dashed lines) lines for the northern (left), southern (center), and western (right) images. The ALMA observations are over-laid on the CO(1--0) (pink) and CO(3--2) (blue) lines from \citetalias{Sharon:2019a} which have been re-scaled to match the same peak flux density as the \ci lines in each image's spectra. The northern image shows a strong difference in line profile shape between the ALMA data and the low-$J$ CO lines.
}
    \label{fig:spectra}
\end{figure*}

We successfully detect both the CO(7--6) and \ci lines at the $>12\,\sigma$ and $>19\,\sigma$ levels, respectively. We show the integrated line maps in Fig.~\ref{fig:intline}, overlaid channel maps in Fig.~\ref{fig:renzos}, and spectra in Fig.~\ref{fig:spectra}. 

There is a clear velocity gradient across the images for both lines, similar to those seen for the CO(1--0) and CO(3--2) lines in \citetalias{Sharon:2019a}. However, the spectral line profiles differ somewhat from those of the lower-$J$ CO lines. While the western image, which is least distorted by gravitational lensing, shows a double-peaked profile similar to those in \citetalias{Sharon:2019a} (as expected for a rotating disk galaxy), the northern image is much more asymmetric in both the CO(7--6) and C\,{\sc i} lines (hence differences between the images' central velocities for the Gaussian fits in Table~\ref{tab:lines}). Since the northern image is actually a merging double image of \emph{part} of J0901, one might expect it to show a more asymmetric profile; however, that was not the case for the two lower-$J$ CO lines in \citetalias{Sharon:2019a}. This difference suggests that there is spatial variation in the CO excitation that results in differential magnification in some of the images. Fitting the combined line profiles to a double Gaussian gives integrated line fluxes of $8.4\pm1.4\,{\rm Jy\,km\,s^{-1}}$ for the CO(7--6) line and $13.4\pm2.0\,{\rm Jy\,km\,s^{-1}}$ for C\,{\sc i}(${\rm ^3P_2\rightarrow ^3\!P_1}$) (including a 10\% flux {scale} calibration uncertainty). We summarize the line fluxes for each image in Table~\ref{tab:lines}.

\begin{deluxetable*}{cllcccc}
\tablewidth{0pt}
\tablecaption{Spectral line profiles and integrated fluxes \label{tab:lines}}
\tablehead{ {Line} & {Parameter} & {Units} & {North} & {South} & {West} & {Total}}
\startdata
CO(7--6) & $S_{\nu,{\rm peak}}$ & ${\rm mJy}$ & $9.2\pm1.6$/$24.4\pm1.8$ & $12.0\pm1.3$/$9.1\pm2.3$ & $5.9\pm0.7$/$5.0\pm0.8$ & $23.1\pm1.8$/$40.7\pm5.1$ \\
{} & FWHM & ${\rm km\,s^{-1}}$ & $121\pm30$/$94\pm10$ & $229\pm24$/$70\pm21$ & $100\pm21$/$106\pm22$ & $192\pm37$/$85\pm10$ \\
{} & $v_{\rm offset}$ & ${\rm km\,s^{-1}}$ & $-76\pm12$/$82\pm4$ & $7\pm16$/$94\pm7$ & $-81\pm8$/$82\pm9$ & $-40\pm18$/$90\pm4$ \\
{} & $S_{\nu}\Delta v$\tablenotemark{*} & ${\rm Jy\,km\,s^{-1}}$ & $3.63\pm0.61$ & $3.61\pm0.63$ & $1.19\pm0.24$ & $8.4\pm1.4$ \\
\hline
C\,{\sc i}(${\rm ^3P_2\rightarrow ^3\!P_1}$) & $S_{\nu,{\rm peak}}$ & ${\rm mJy}$ & $12.4\pm0.8$/$31.8\pm2.2$ & $21.4\pm1.3$/$10.1\pm2.9$ & $9.1\pm0.8$/$7.5\pm1.0$ & $33.4\pm2.3$/$58.7\pm5.3$ \\
{} & FWHM & ${\rm km\,s^{-1}}$ & $143\pm34$/$102\pm11$ & $222\pm17$/$55\pm18$ & $149\pm28$/$126\pm29$ & $178\pm31$/$114\pm11$ \\
{} & $v_{\rm offset}$ & ${\rm km\,s^{-1}}$ & $-89\pm13$/$83\pm4$ & $23\pm9$/$55\pm18$ & $-84\pm12$/$80\pm13$ & $-65\pm16$/$86\pm5$ \\
{} & $S_{\nu}\Delta v$\tablenotemark{*}& ${\rm Jy\,km\,s^{-1}}$ & $5.32\pm0.86$ & $5.65\pm0.78$ & $2.44\pm0.46$ & $13.4\pm2.0$ \\
\enddata
\tablenotetext{*}{The integrated line flux includes both peaks, the statistical uncertainty, and an additional 10\% flux calibration uncertainty.}
\tablecomments{Line profiles are fits to two Gaussians with centroid velocity offsets measured relative to the $z=2.2586$ ${\rm H\alpha}$ systemic redshift from \citet{Hainline:2009a}. The parameters for each Gaussian are separated with slashes in the table.}
\end{deluxetable*}

\section{Analysis}\label{sec:analysis}

\subsection{Radio {Spectrum} and the Decomposition of Radio Emission}\label{subsec:radio_decomposition}

Now that we have obtained the {${1.4\,{\rm GHz}}$ and ${3\,{\rm GHz}}$} continuum maps, we can disentangle the radio continuum emission mechanisms. The cross-band spectral index is calculated using the expression $\alpha_{\rm a}^{\rm b} =  {\rm log} (S_{\rm a}/S_{\rm b}) /  {\rm log}(a/b) $, and we find the spectral index between 1.4\,GHz and 3\,GHz bands (observed frequencies; corresponding to $4.6\,{\rm GHz}$ and $9.8\,{\rm GHz}$ in the rest frame) is $\alpha_{4.6}^{9.8} \sim -0.54$. This value is {a bit flatter than that of} star-forming galaxies {in the local Universe \citep[e.\/g.\/,][]{Condon:1992a} and at higher redshift \citep[e.\/g.\/,][]{Algera:2021a, Algera:2022a}, and is closer to that of individual star-forming regions in local galaxies \citep[e.\/g.\/,][]{Linden:2020a} and $z\sim1.2$ galaxies \citep[e.\/g.\/,][]{Murphy:2017a}. The relatively flat spectral index suggests that} the higher frequency (e.\/g.\/, $>10$\,GHz) radio emission is not severely contaminated by AGN-driven synchrotron emission{, and may have a significant contribution from free-free emission}. 

We {therefore} combine these VLA continuum flux {density} measurements with the Ka-band {(${\nu_{\rm obs}=35\,{\rm GHz}}$}) continuum emission detected in \citetalias{Sharon:2019a} in order to constrain the contributions of the free-free (FF) and synchrotron emission to the radio spectrum. The Ka-band data have low SNR, with the northern and southern arcs {are} only weakly detected and the western arc {is} not detected. We therefore use the sum of the two detected images' peak flux density values for the total 35\,GHz flux density. This choice results in a total flux density of $S_{\rm 35\,GHz}=233.8\pm44.3\rm\ \mu Jy$. The  {${1.4\,{\rm GHz}}$ and ${3\,{\rm GHz}}$} emission, however, are significantly detected across multiple emission elements. Therefore, in order to make a fair comparison between the three bands, we (1) smooth the {${1.4\,{\rm GHz}}$ and ${3\,{\rm GHz}}$} maps to match the spatial resolution of the Ka-band images, and (2) add noise to the {${1.4\,{\rm GHz}}$ and ${3\,{\rm GHz}}$} maps such that the SNR of the brightest peak matches that of the {${35\,{\rm GHz}}$} images. The former step ensures that nearby features within J0901 are blended at the same level across frequencies. The latter step ensures that we are similarly insensitive to low surface brightness extended emission in all three bands. While this somewhat conservative choice does inflate our uncertainties in the radio spectrum decomposition, it more accurately captures the total radio emission from J0901 when we need to use different flux {density} extraction methods for different frequencies \footnote{Using a smaller matched aperture based on the  {${35\,{\rm GHz}}$} emission would only measure the flux {density} from a small part of J0901 (rather than the whole galaxy's radio emission), while still inflating the noise, but for the  {${35\,{\rm GHz}}$} flux {density} instead. Using a larger matched aperture based on the extended emission in the  {${1.4\,{\rm GHz}}$} and  {${3\,{\rm GHz}}$} maps then captures the  {${35\,{\rm GHz}}$} noise peaks that produce the un-physically large {flux densities} seen in \citepalias{Sharon:2019a} that we have ruled out based on the lower-frequency observations.}. 
To match the observations in these ways, we first convolve the  {${1.4\,{\rm GHz}}$ and ${3\,{\rm GHz}}$ flux density} maps with 2D Gaussians that deliver the 35\,GHz resolution ($2.\arcsec2\times 2.\arcsec0$). We then add random noise (convolved with the beam) such that the peak SNR for the three bands is the same. The noise added through this strategy is summarised in Table~\ref{tab:addnoise_conv}. 

\floattable
\begin{deluxetable*}{cccc}
\tablewidth{0pt}
\tablecaption{Additional noise required to match the 35\,GHz map resolution and peak SNR
 \label{tab:addnoise_conv}}
\tablehead{ {Map} & {Peak Flux Density} & {RMS Noise} & {Noise Required} \\
{$\nu_{\rm obs}$} & ($\mu\rm Jy\,beam^{-1}$) & ($\mu\rm Jy\,beam^{-1}$) & ($\mu\rm Jy\,beam^{-1}$) 
}
\startdata
 {${1.4\,{\rm GHz}}$} & 72.8 & 2.0 & 11.0\\
 {${3\,{\rm GHz}}$} & 45.4 & 1.7 & 6.4\\
 {${35\,{\rm GHz}}$} & 148.5  & 26.6 & {-}\\
\hline
\enddata
\tablecomments{The $1.4\,{\rm GHz}$ and $3\,{\rm GHz}$ maps have been convolved to the $35\,{\rm GHz}$ beam size. Therefore, the flux densities and RMS are different from their values in Table~\ref{tab:maps_flux}.}
\end{deluxetable*}

\begin{figure*} 
	\centering
	\vspace{0cm}
	\subfigtopskip=0pt 
	\subfigbottomskip=0pt 
	\subfigcapskip=0pt
	\includegraphics[width=0.95\linewidth]{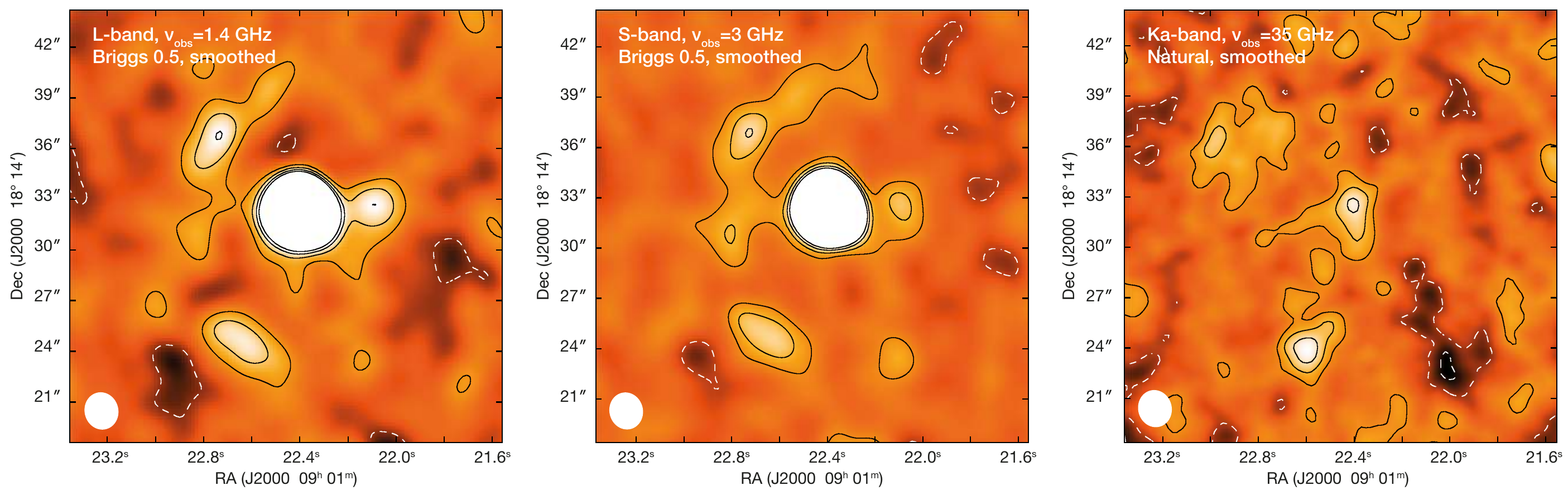}
	\caption{Briggs0.5 weighted continuum flux {density} maps from VLA  {${1.4\,{\rm GHz}}$} (left) and  {${3\,{\rm GHz}}$} (center), after matching to the peak sensitivity and resolution of the VLA 35\,GHz continuum map (right). The synthesized beam is $2.\arcsec2\times 2.\arcsec0$, shown in white in the lower-left corners. For the {${1.4\,{\rm GHz}}$} data, Gaussian noise of $11.01\,\mu\rm Jy\ beam^{-1}$ is added, and for the {${3\,{\rm GHz}}$} data $6.44\,\mu\rm Jy\ beam^{-1} $ of Gaussian noise is added. The additional noise results in a matched peak SNR for J0901 of 5.6 in all three bands, which minimizes the contribution of extended emission revealed by the maps' different sensitivities to the overall flux  {density} calculations. Contour levels are $(-2,2,4,6,8)\times$RMS for the  {${1.4\,{\rm GHz}}$} and ${3\,{\rm GHz}}$ maps, and $(-3,-1.5,1.5,3,4.5)\times$RMS for the $35\,{\rm GHz}$ {map}; the maps resulting noise levels are listed in Table~\ref{tab:addnoise_conv}. Positive and negative contours are distinguished with black (solid) and white (dashed) colors, respectively. We set a maximum threshold of 80 $\mu\rm Jy\,beam^{-1}$ for the  {${1.4\,{\rm GHz}}$ map} and 60 $\mu\rm Jy\,beam^{-1}$ for the  {${3\,{\rm GHz}}$ map} in color scale. The central bright lens is therefore ignored, and the J0901 flux  {density} distribution is better seen.}
	\label{fig:smth-flux_maps}
\end{figure*}

These matched resolution/peak-SNR VLA maps are shown in Fig.~\ref{fig:smth-flux_maps}. Because of the lower resolution, the multiple peaks detected in the  {${3\,{\rm GHz}}$} southern and western arcs are no longer distinguishable. At the $2\,\sigma$ level, the central lens and western arc are smeared together. We give the updated flux  {density measurements after ${1.4\,{\rm GHz}}$ and ${3\,{\rm GHz}}$} are matched to the 35\,GHz resolution and peak sensitivity in Table~\ref{tab:convolved_flux}. Since the emission is still noticeably extended even when the data are smoothed and noise has been added, we still extract the flux {densities} using an aperture in these two bands. The integrated fluxes {densities} do not change significantly after this matching process.

\floattable
\begin{deluxetable*}{cccccc}
\tablewidth{0pt}
\tablecaption{J0901 continuum flux densities used for the radio spectrum decomposition
\label{tab:convolved_flux}}
\tablehead{ {Map} & {RMS} & {North} & {South} & {West} & {Total}\\
{$\nu_{\rm obs}$} & {$(\rm\mu Jy\, beam^{-1})$} & {$(\rm\mu Jy)$} &{$(\rm\mu Jy)$} &{$(\rm\mu Jy)$} & {$(\rm\mu Jy)$}  
} 
\startdata
 {${1.4\,{\rm GHz}}$} & 13.0 &  {${267\pm42}$} &  {${140\pm32}$} &  {${117\pm23}$} &  {${524\pm61}$}\\
 {${3\,{\rm GHz}}$} & 7.8  &  {${193\pm26}$} &  {${98\pm19}$} &  {${51\pm13}$} &  {${342\pm37}$}\\
 {${35\,{\rm GHz}}$} & 26.6 & $85\pm28$ & $149\pm31$ & {-} & $234\pm44$\\
\enddata
\tablecomments{The $1.4\,{\rm GHz}$ and $3\,{\rm GHz}$ flux densities are integrated values after matching resolution and peak SNR to the $35\,{\rm GHz}$ map. Flux density at 35\,GHz is derived from the peak pixel values; we ignore the western arc for the $35\,{\rm GHz}$ flux density measurement because the SNR is below unity.}
\end{deluxetable*}

Using these matched resolution/peak-SNR {flux densities}, we fit the emission to FF and non-thermal components using the methods in \citet{Algera:2021a}. The  {${1.4\,{\rm GHz}}$, ${3\,{\rm GHz}}$, and ${35\,{\rm GHz}}$ observed} frequencies correspond to rest-frame 4.6\,GHz, 9.8\,GHz, and 114.1\,GHz respectively.  {While this} frequency range should be dominated by FF and non-thermal emission, we note that the  {${\nu_{\rm obs}=35\,{\rm GHz}}$ data} may be partly contaminated by modified black body (MBB) emission from dust.  {Although significant dust contribution to the ${35\,{\rm GHz}}$ flux density is unlikely based on the spectral index between it and the ${1.4\,{\rm GHz}}$ data (${\alpha_{9.8}^{35}=\sim-0.3}$, indicative of free-free emission),} we employ the modified blackbody (MBB) fitting procedure outlined in \citet{Algera:2023a} {in order to more precisely evaluate the potential contribution}. We incorporate the IR measurements from \citet{Saintonge:2013a}, excluding the two shortest wavelengths (70\,$\rm \mu m$ and 100\,$\rm \mu m$), which are dominated by hot dust emission. We also substitute the 1.2\,mm flux density with the ALMA  {band} 6 measurement presented here. In addition, we include the ALMA Band 3 {(${\lambda_{\rm obs}=3\,{\rm mm}}$)} continuum flux density from \citet{Liu:2023a}, which yields $S_{\rm 3\,mm}=\rm 1.47\pm 0.19\,m Jy$. Assuming optically thick emission for $\rm \lambda < 100 \mu m$ and treating the emissivity ($\beta$) as a free parameter, the predicted contribution to the  {${35\,{\rm GHz}}$} flux density from cold dust emission is $25.2_{-3.8}^{+4.5}\rm \mu Jy$. We obtain similar results (20.9 $\rm \mu Jy$) when using the IR fitting tool \texttt{cmcirsed} \citep{Casey:2012a} on the same data. Since this flux density is below the noise level of the  {${35\,{\rm GHz}}$} continuum emission, we ignore its contribution when fitting the radio spectrum and assume that the radio emission across the three bands consists of only FF and non-thermal emission.

{For the continuum fit, we} assume that both the thermal and non-thermal emission have power law forms \citep[e.\/g.\/,][]{Condon:1992a, Murphy:2017a, Tabatabaei:2017a, Algera:2021a}. The decomposition of the two components can then be expressed as

\begin{equation}\label{eq:radio_decomposition}
S_\nu = \left(1-f_{th}\right)S_{\nu_0}\left(\frac{\nu}{\nu_0}\right)^{\rm\alpha_{NT}}+f_{th}S_{\nu_0}\left(\frac{\nu}{\nu_0}\right)^{-0.1}.
\end{equation}

\noindent
Following the process of \citet{Algera:2021a}, the reference frequency $\nu_0$ is arbitrarily set to the  {\emph{observed-frame}} 1.4\,GHz, which is also where the reference flux {density} is given ($S_{\nu_0}$). The FF emission spectral index is fixed at $-0.10$ as is common practice. The $\rm\alpha^{NT}$ is the non-thermal spectral index, and $f_{th}$ is the FF emission fraction measured at $\nu_0$. There are thus three free parameters to be constrained: $S_{\nu_0}$, $f_{th}$ and $\rm\alpha^{NT}$.

\begin{figure*} 
	\centering
	\vspace{0cm}
	\subfigtopskip=0pt 
	\subfigbottomskip=0pt 
	\subfigcapskip=0pt
	\subfigure{
		\label{fig:mcmc-flat}
		\includegraphics[width=0.49\linewidth]{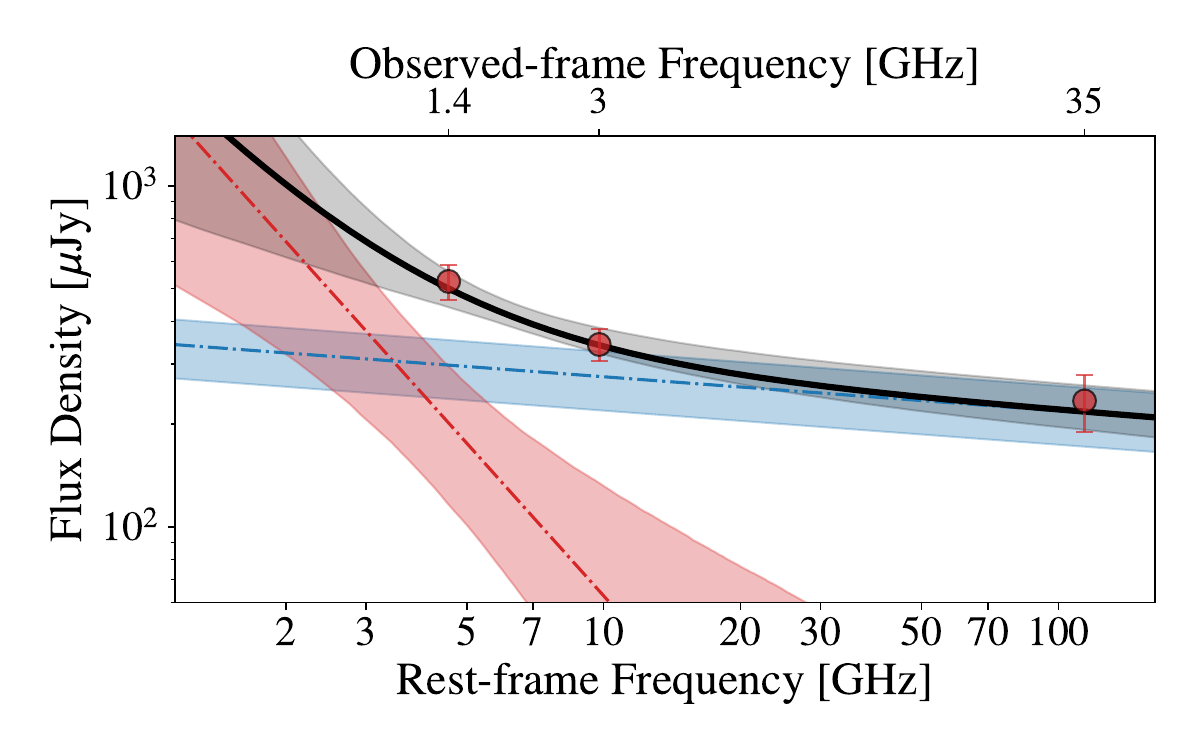}}
	\subfigure{
		\label{fig:mcmc}
		\includegraphics[width=0.49\linewidth]{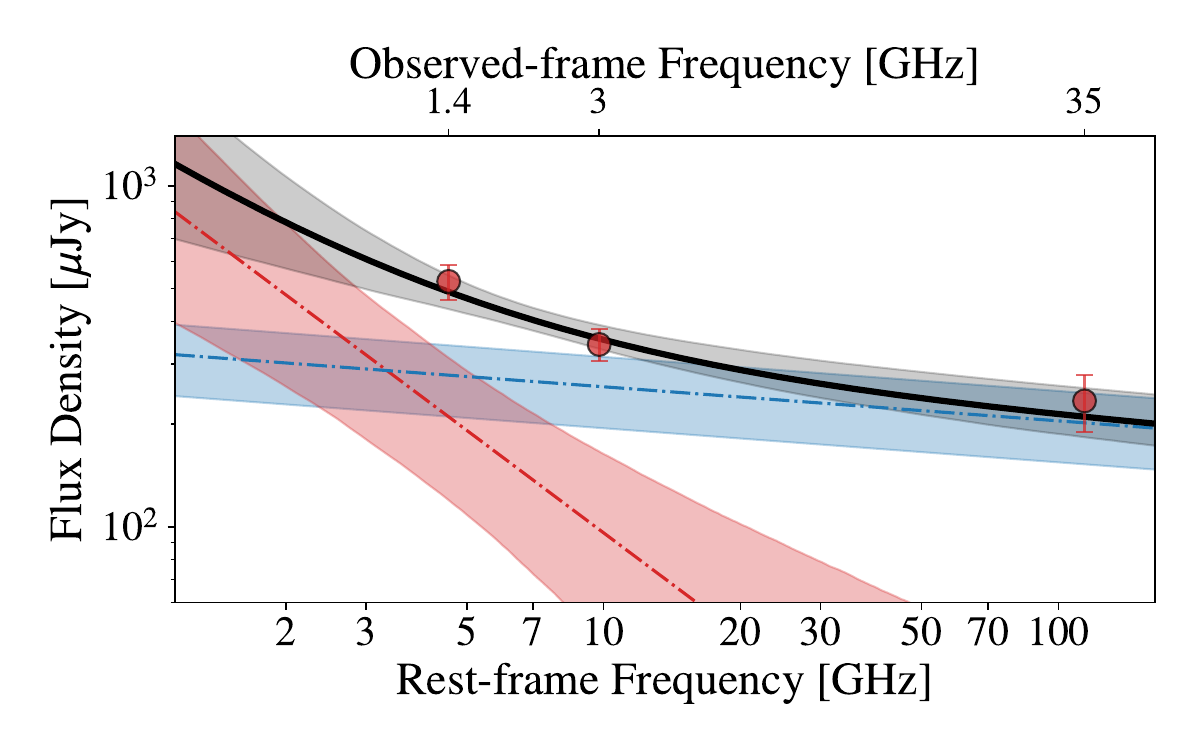}}
	\caption{Best-fit radio spectra assuming a flat prior for the non-thermal spectral index (left) or a Gaussian prior for the non-thermal spectral index (right). Red points with uncertainties are the three VLA measurements. The solid black line is the best-fit model, with the contributions from the FF and non-thermal emission shown in blue and red dashed lines, respectively. The shaded regions show the 1\,$\sigma$ confidence regions for the lines of the corresponding colors.}
	\label{fig:mcmc-2plots}
\end{figure*}

\begin{figure*} 
	\centering
	\vspace{0cm}
	\subfigtopskip=0pt 
	\subfigbottomskip=0pt 
	\subfigcapskip=0pt
	\subfigure{
		\label{fig:corner-flat}
		\includegraphics[width=0.49\linewidth]{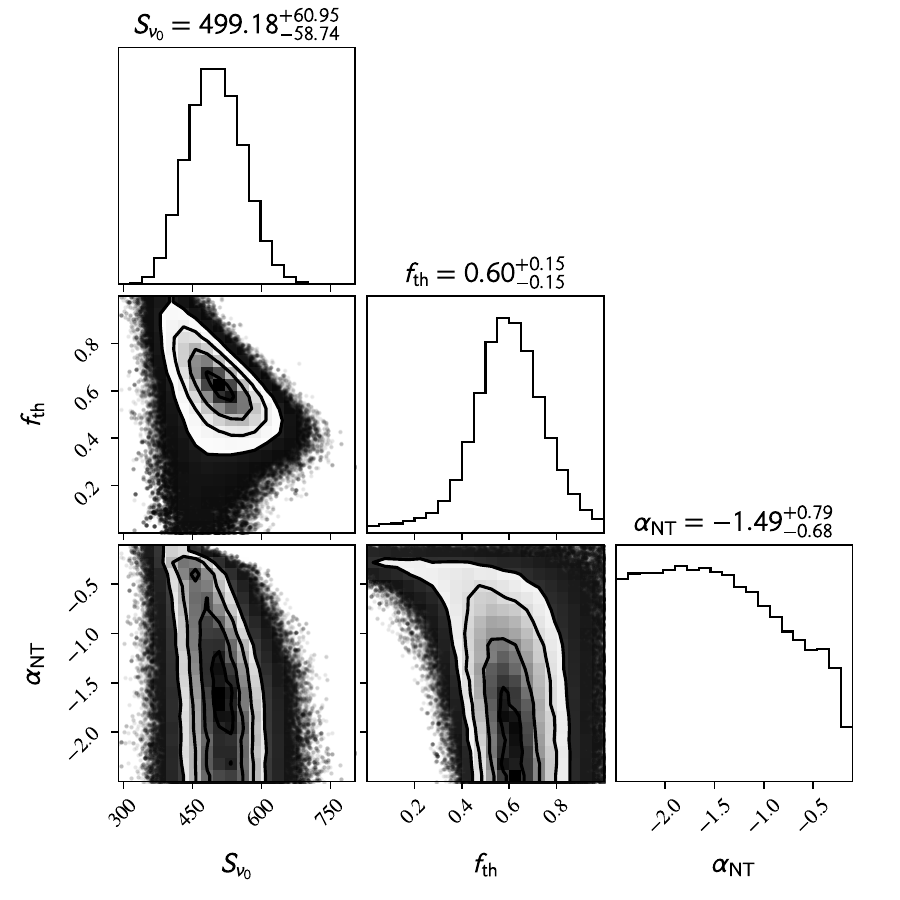}}
	\subfigure{
		\label{fig:corner}
		\includegraphics[width=0.49\linewidth]{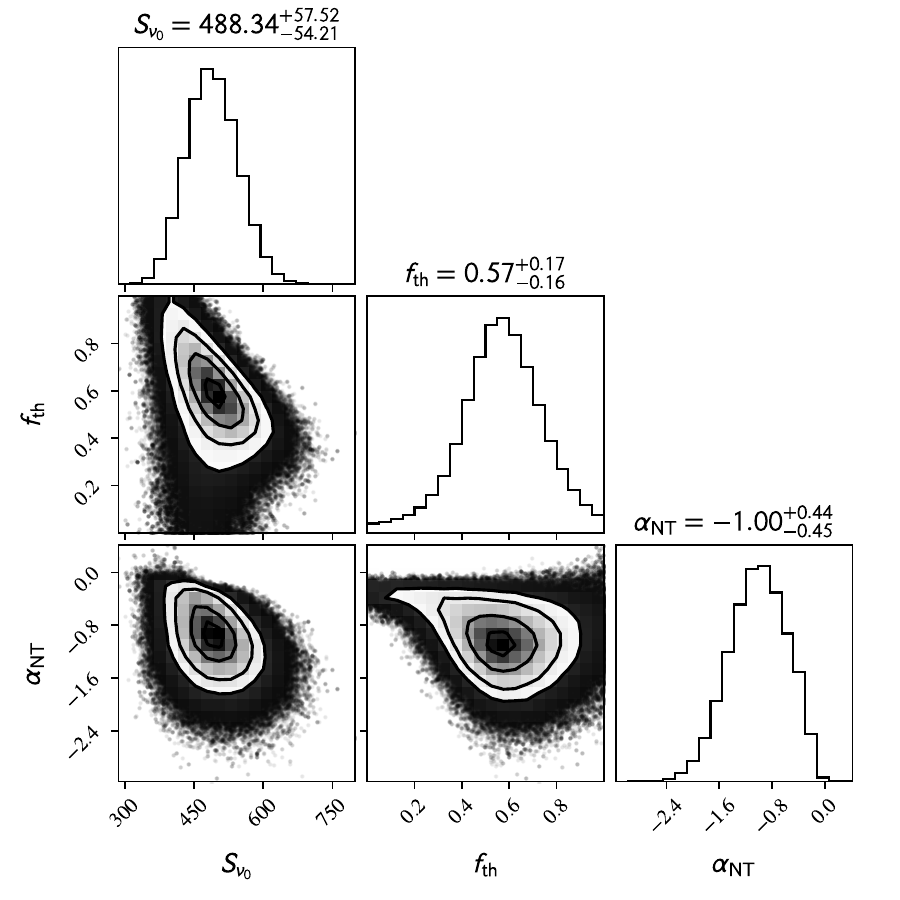}}
	\caption{The corner plot (marginalized 1D and 2D probability distributions) for the MCMC analysis of the radio spectrum decomposition and best-fit values. Left: MCMC fit results with a flat prior assumption for the non-thermal spectral index. Right: MCMC fit results with a Gaussian prior assumed for non-thermal spectral index.}
	\label{fig:corner-2plots}
\end{figure*}

We run a Markov chain Monte Carlo (MCMC) analysis to constrain the three fit parameters for the radio SED. We adopt a flat prior on the thermal fraction for $0\leq f_{th} \leq 1$ throughout. We first attempt to fit without an initial guess for $\rm\alpha^{NT}$, but rather constrain it within a range between $-2.5$ and $-0.1$. We show the best-fit spectrum in Fig.~\ref{fig:mcmc-flat} and the MCMC corner plots in Fig.~\ref{fig:corner-flat}. In this case, $\rm\alpha^{NT}$ is poorly constrained,  {so we do not ascribe much significance to the rather steep spectral index (${{\rm \alpha^{NT}=-1.49^{+0.79}_{-0.68}}}$)}. However, the reference flux {density} $S_{\nu_0}$ and the thermal fraction $f_{th}$ are {better constrained. While the fit indicates} that the three bands' observations are all dominated FF emission{, the degree of agreement or disagreement with thermal fractions found for other galaxies cannot be accurately assessed; we would need to use the poorly constrained non-thermal spectral index to extrapolate to frequencies where the thermal fraction is commonly calculated, and the spectral index would introduce too much uncertainty}.

Given the lack of observational constraints in the region of the spectrum more likely to be dominated by non-thermal emission, {and the poorly constrained and unusually steep non-thermal index,} we also re-run the MCMC analysis with priors on the non-thermal radio spectral index.  {In the context of the Local Universe, the typical synchrotron power-law slope is ${{\rm \alpha^{NT}\sim -0.85}}$ \citep{Niklas:1997a, Murphy:2011a}. The scatter is found to be 0.3-0.5 in large radio samples \citep{Calistro:2017a, Smolcic:2017a, Gim:2019a}. We therefore} assume that a majority of the non-thermal emission is synchrotron and set a Gaussian prior on $\rm{\alpha^{NT}}$ with a mean of $-0.85$ and a standard deviation of 0.5, following \citet{Algera:2021a}. We show the resulting best-fit spectrum with this Gaussian prior in Fig.~\ref{fig:mcmc}, and the corner plots of the MCMC in Fig.~\ref{fig:corner}. In this case, the best-fit  {${{\rm \alpha^{NT}=-1.00^{+0.44}_{-0.45}}}$} is somewhat flatter and better constrained than the previous best-fit value (as seen when adding priors in \citealt{Westcott:2018a}), but it still has substantial uncertainties making it technically consistent with expected indices at low and high redshift (${\alpha^{\rm NT}\sim-0.8}$; \citealt{Condon:1992a,Algera:2022a}). If scaled to the rest-frame 1.4\,GHz, the thermal fraction $f_{th}$ is {${0.31_{-0.19}^{+0.31}}$. This value is high relative to the typical value of ${\sim0.1}$ found} for star-forming galaxies  {at both low and high redshifts \citep[e.\/g.\/.,][]{Condon:1992a, Algera:2021a}, but is technically consistent due to the large uncertainty (mostly driven by the non-thermal spectral index)}. Improved constraints on the thermal fraction could be obtained both through higher SNR {${35\,{\rm GHz}}$} continuum measurements, and additional measurements nearer to (and at lower frequencies than) 1.4\,GHz{; the latter would be particularly important for better constraining the non-thermal spectral index, which in turn affect all extrapolated thermal fractions}.

\subsection{The {Infrared-Radio} Correlation}\label{subsec:IRRC}

Since J0901 is known to contain an AGN \citep{Fadely:2010a, Genzel:2014a}, it is important to establish the degree to which the radio continuum emission may be contaminated by the AGN prior to using it as an alternative SFR tracer (Sections~\ref{subsec:sfr} and \ref{subsec:skrelation}). While there  {is} no sign of a particularly strong non-thermal/synchrotron component in the radio  {spectrum} (Section~\ref{subsec:radio_decomposition}),  {the radio spectrum is not very well sampled and the best-fit non-thermal spectral index had significant uncertainty}. 

It is known that the radio continuum and infrared have a tight correlation {(the IRRC)} among star-forming {galaxies \citep[e.\/g.\/,][]{Helou:1985a, deJong:1985a}}, usually expressed as

\begin{equation}\label{eq:q}
q_{\rm TIR}={\rm log}\left( \frac{L_{\rm TIR}}{3.75\times10^{12}{\rm W}} \right) - {\rm log}\left( \frac{L_{\rm 1.4\,GHz}}{\rm W\ Hz^{-1}}  \right), 
\end{equation}

\noindent
which depends on the ratio between the total infrared ($8$--$1000\,{\rm \mu m}$) and 1.4\,GHz luminosities. Adopting the Gaussian prior MCMC result from Section~\ref{subsec:radio_decomposition}, we obtain a rest-frame 1.4\,GHz flux density of {${S_{\rm 1.4\,GHz, rest}=998_{-417}^{+1041}\,{\rm \mu Jy}}$} (which is not corrected for magnification). We adopt the total infrared (TIR) luminosity from \citet{Saintonge:2013a} {to obtain ${L_{\rm TIR}=1.8\times 10^{12}(\mu/30)\,L_\odot}$, where ${\mu=30}$ is} the typical magnification {from \citetalias{Sharon:2019a}}. These two quantities combined yield {${q_{\rm TIR}=2.65_{-0.31}^{+0.24}}$} for J0901.

From the  {IRRC} derived in the Local Universe, a normal star-forming galaxy is expected to have a  {IRRC} around $\sim 2.64$ \citep{Bell:2003a}. Our estimate of $q_{\rm TIR}$ suggests that J0901 is a normal star-forming galaxy with very limited contamination from AGN in the radio.  {While predictions have varied on whether and by how much} $q_{\rm TIR}$ {should change with redshift \cite[e.\/g.\/,][]{Carilli:2001a, Murphy:2009a, Lacki:2010a, Lacki:2010b, Schober:2016a}, observations have shown that high-redshift star-forming galaxies tend} to have lower  {$q_{\rm TIR}$}. For example, \citet{Delhaize:2017a} find an evolving $q_{\rm TIR}$ of the form $q_{\rm TIR}=(2.88\pm 0.03)(1+z)^{-0.19\pm 0.01}$. From this equation, the expected $q_{\rm TIR}$ at the redshift of J0901 ($z=2.26$) is $2.3\pm 0.05$. Likewise, \citet{Algera:2020a} conclude that high-redshift dusty star-forming galaxies {(DSFGs)} tend to have $q_{\rm TIR}$ $\sim0.4$ dex lower than that in the Local Universe. \citet{Delvecchio:2021a} discovered that the correlation primarily evolves with stellar mass and the redshift dependence is secondary. The expected $q_{\rm TIR}$ is about 2.43 considering the stellar mass {of} J0901{.}

Another method for quantifying the radio excess of galaxies is the $r$ parameter, defined as  {${r={\rm log}( L_{\rm 1.4\,GHz, rest}/{\rm SFR_{IR}})}$} \citep[as used in, e.\/g.\/,][]{Novak:2017a, Delvecchio:2017a}.  Galaxies with $r>22\times (1+z)^{0.013}$ are believed to have some contribution from AGN in the radio. For J0901, correcting for an identical $\mu=30$ at both wavelengths,  {${r=21.2_{-0.2}^{+0.3}}$} which is below the redshift-appropriate threshold of $r=22.3${.}

{Based on these results, J0901 has a ${q_{\rm TIR}}$ (or ${r}$) at the lower edge of the expected range for star-forming galaxies at its redshift \citep{Delhaize:2017a}. The moderate offset is well within the uncertainty of the measured values for J0901, through it is possible that processes like differential lensing \citep[e.\/g.\/,][]{Blandford:1992a} or limitations of the SED fitting \citep[e.\/g.\/,][]{Algera:2020a} may affect these results.} We therefore conclude that it is unlikely that the AGN in J0901 is a significant source of radio emission, and thus that the radio continuum likely originates in processes related to star formation. The excess  {${35\,{\rm GHz}}$ emission} noted by \citepalias{Sharon:2019a} is more likely due to the low SNR of the detection and the resulting ambiguity in which emission peaks can be ascribed to J0901 at the poorer resolution of those observations. 

\subsection{Radio-derived Star Formation Rates}\label{subsec:sfr}

Having established that the AGN in J0901 is unlikely to be a significant source of radio continuum emission, we can use our new observations to better establish the galaxy's SFR. While multiple SFR tracers have been observed for J0901 previously, there is not perfect agreement between these measurements even when all are corrected for lensing using the same model. For example, H$\alpha$ is considered a robust tracer because it is directly related to the newly-formed massive stars, and is thus sensitive to star formation on timescales $\lesssim\rm 10\,Myr$. However, H$\alpha$ can suffer from significant extinction due to dust in the star-forming environment. \citetalias{Sharon:2019a} found a $\rm H\alpha$-derived SFR of $14.5_{-1.7}^{+2.1}\,M_\odot\,{\rm yr}^{-1}$ when not correcting for extinction. However, using the H$\alpha$/H$\beta$ ratio measured in \citet{Hainline:2009a} to correct for extinction, the $\rm SFR_{H\alpha}$ will increase by a factor of $\gtrsim$ 20 (albeit with considerable uncertainty due to the fractional coverage of J0901's extended line emission by the slit spectroscopy). Given the significant dust obscuration in J0901, the TIR may be a better SFR tracer (sensitive to star formation within $\lesssim\rm 100\,Myr$). Using the \citet{Saintonge:2013a} TIR, corrected for the magnification as in \citetalias{Sharon:2019a}, $\rm SFR_{\rm TIR}=268_{-61}^{+63}\, M_\odot\,\rm yr^{-1}$. Another SFR probe is the combination of the H$\alpha$ and TIR emission as per \citet{Kennicutt:2012a}. This hybrid tracer gives a much lower SFR of ${\rm SFR_{H\alpha+TIR}}=103_{-20}^{+21}\,M_\odot\,\rm yr^{-1}$ (albeit comparable to the extinction-corrected H$\alpha$-derived SFR using the H$\alpha$/H$\beta$ ratio). However, we note that this hybrid tracer relation is derived from a local Universe sample with ${\rm log}\left(L_{\rm TIR}/L_\odot\right) < 11.9$ and therefore may not be valid for extremely dust-luminous systems in the early Universe like J0901. 

As radio continuum is an extinction-free SFR tracer, our new VLA observations may help better-determine the SFR for J0901. Since we decomposed the FF and non-thermal components of the radio emission in Section~\ref{subsec:radio_decomposition}, we can  {calculate and compare} the radio-derived SFR from these {two} components following \citet{Murphy:2012a}. The FF luminosity is proportional to the production rate of ionizing photons in {H\,{\sc ii}} regions, and therefore sensitive to current ($\sim$ 10\,Myr lifetimes of very massive stars) SFR. We can determine that SFR using the expression {from \citet{Murphy:2012a}}

\begin{equation}\label{eq:SFR_th}
\begin{aligned}
\frac{\rm SFR_\nu^T}{M_\odot\,{\rm yr}^{-1}} &= 4.6\times 10^{-28}\left(\frac{T_e}{10^4\,{\rm K}}\right)^{-0.45}\left(\frac{\nu}{\rm GHz}\right)^{0.1}\\
&\ \times {\frac{L_\nu^T}{\rm ergs\,s^{-1}\,Hz^{-1}}}
\end{aligned}
\end{equation}

\noindent
where $T_e$ is the electron temperature\footnote{Note that the superscripts on the SFRs do not indicate some fraction of the true/total SFR; the superscripts just indicate which expression/emission component is used to derive the SFR.}. The ionizing photon production rate (and hence SFR) has a weak dependence on the electron temperature \citep{Rubin:1968a}, which we assume to be $\rm 10^4\,K$. The non-thermal component comes from synchrotron emission associated with the cosmic rays (CR) from supernova remnants, and hence traces a delayed SFR on a slightly longer timescale \citep[$\sim$30--100\,Myr lifetimes for massive stars and radiating CR electrons;][]{Bressan:2002a}. The SFR inferred from the non-thermal synchrotron emission is based on the total core-collapse supernova rate, and is given by 

\begin{equation}\label{eq:SFR_nth}
\frac{\rm SFR_\nu^{NT}}{M_\odot\,{\rm yr}^{-1}}=6.64\times 10^{-29}\left(\frac{\nu}{\rm GHz}\right)^{\alpha^{NT}}{\frac{L_\nu^{NT}}{\rm ergs\,s^{-1}\,Hz^{-1}}}
\end{equation}

\noindent
{from \citet{Murphy:2012a}} where $\rm\alpha^{NT}$ is the assumed spectral index (constrained by our radio SED).
Since the SFR produces both the FF and non-thermal emission, {one} can invert Equations~\ref{eq:SFR_th} and ~\ref{eq:SFR_nth} and sum them to obtain the expected total luminosity at some frequency ($L_\nu$) as function of the SFR. One can then use that expression to determine the SFR based on a measured $L_\nu$, which is given by

\begin{equation}\label{eq:SFR_combine}
\begin{aligned}
\frac{\rm SFR_\nu^{T+NT}}{M_\odot\,{\rm yr}^{-1}} &= \rm 10^{-27}\Big [ 2.18 \left(\frac{T_e}{10^4 {\rm K}}\right)^{0.45} \left(\frac{\nu}{\rm GHz}\right)^{-0.1} \\
&\ + 15.1\left(\frac{\nu}{\rm GHz}\right)^{-\alpha^{NT}}\Big ]^{-1} \frac{L_\nu}{\rm ergs\,s^{-1}\,Hz^{-1}}
\end{aligned}
\end{equation}

\noindent
{from \citet{Murphy:2012a}.} Given the mismatch in SFR timescales for the FF and non-thermal emission, this expression is valid when star formation is approximately constant for $\gtrsim30\,{\rm Myr}$. In all three of the above equations, $\nu$ is the rest-frame frequency where the SFR is calculated (set to 1.4\,GHz). $\rm L_\nu^T$, $\rm L_\nu^{NT}$ and $\rm L_\nu$ are the FF, non-thermal, and total luminosities at $\rm\nu$.

Using the radio continuum decomposition with the Gaussian prior on the non-thermal spectral index, the rest-frame 1.4\,GHz FF and non-thermal emission are 310{$_{-110}^{+140}\,\mu$Jy and 680$_{-470}^{+1100}\,\mu$}Jy, respectively. We obtain the SFRs as ${\rm SFR^T}={620}_{-220}^{+280}\,M_\odot\,\rm{yr}^{-1}$, $\rm{ SFR^{NT}}={230}_{-160}^{+570}\,M_\odot\,{\rm yr}^{-1}$, and $\rm{SFR^{T+NT}}={280}_{-120}^{+460}\,M_\odot\,{\rm yr}^{-1}$ (as listed in Table~\ref{tab:sfr}). These SFRs are all consistent with TIR-derived SFR given the uncertainties except for the FF emission. The non-thermal emission derived SFR is also consistent with the H$\alpha$+TIR hybrid SFR. Despite the considerable uncertainties, the radio-derived SFRs do suggest a significant fraction of the SFR in J0901 is dust-obscured, and the true SFR is much higher than inferred from the H$\alpha$ luminosity (uncorrected for extinction) alone.

\begin{deluxetable}{lcl}
\tablewidth{0pt}
\tablecaption{Radio-derived SFRs
\label{tab:sfr}}
\tablehead{ {Tracer} & {SFR$_{\rm GP}$}\tablenotemark{a} & {SFR$_{\rm flat}$}\tablenotemark{b} \\
{} & {$(M_\odot\rm\,yr^{-1})$} & {$(M_\odot\rm\,yr^{-1})$} 
} 
\startdata
Free-free &  {${{620}_{-220}^{+280}}$} &  {${{670_{-230}^{+270}}}$} \\
Non-thermal &  {${{230}_{-160}^{+570}}$} &  {${{410}_{-320}^{+1880}}$} \\
Combined &  {${{280}_{-120}^{+460}}$} &  {${{440}_{-260}^{+1450}}$} \\
\enddata
\tablenotetext{a}{From Gaussian prior MCMC fit.}
\tablenotetext{b}{From flat assumption MCMC fit.}
\end{deluxetable}

\subsection{C\,{\sc i} and 1.2\,mm derived gas masses}\label{subsec:gasmass}

Given the substantial uncertainty in CO-to-${\rm H_2}$ conversion factors, it is worth testing alternative probes of the gas mass for consistency. \citetalias{Sharon:2019a} found that J0901 has a CO(1--0)-derived gas mass of $M_{gas}=(1.6^{+0.3}_{-0.2})\times10^{11}(\alpha_{CO}/4.6)\,M_\sun$. However, {their} rest-frame $877\,{\rm \mu m}$ continuum observations {from the Submillimeter Array (SMA)} gave gas masses in the range of ($\sim5$--7)$\times10^{10}\,M_\sun$ {using the methods from \citet{Scoville:2016a}} (depending on temperature assumptions), which suggested a lower, more U/LIRG-like $\alpha_{CO}$ might be more appropriate for J0901. Here we can use the \ci and ALMA continuum detections as alternative probes of the gas mass. 

Given the integrated total \ci flux of $13.4\pm2.0\,{\rm Jy\,km\,s^{-1}}$, we can  {calculate} a neutral carbon mass using the equation

\begin{equation}\label{eq:ci_gas_mass}
    M_{\rm C\,{\textsc i}} = 4.56\times10^{-3}Q(T_{ex})\frac{1}{5}e^{62.5/T_{ex}}L^\prime_{\rm C{\textsc i}(2-1)}M_\odot
\end{equation}

\noindent 
from \citet{Weiss:2003a}, where $Q(T_{ex})=1+3e^{-23.6/T_{ex}}+5e^{-62.5/T_{ex}}$. This expression assumes that the C\,{\sc i} emission is optically thin and that the gas is in local thermodynamic equilibrium (LTE), which may not be the case for all dusty high-redshift galaxies like J0901 \citep[e.\/g.\/,][]{Glover:2015a, Harrington:2021a, Papadopoulos:2022a}. \citet{Papadopoulos:2022a} in particular find that \ci has highly subthermal excitation in a sample of 106 galaxies. The unknown excitation of J0901 therefore brings additional systematic uncertainties into our analysis. Equation~\ref{eq:ci_gas_mass} also requires an assumption of $T_{ex}$, since we have not measured both C\,{\sc i} fine structure lines. We assume $T_{ex}=30\,{\rm K}$, based on the typical values found in \citet{Harrington:2021a} and \citet{Walter:2011a}, and similarity to the $T_{\rm dust}=36\,{\rm K}$ for J0901 from \citet{Saintonge:2013a}. Given these assumptions, we find ${M_{\rm C{\textsc i}}=(4.96\pm0.75)\times10^7 (\mu/30)\, M_\sun}$ where $\mu\approx30$ is the typical magnification factor found in \citetalias{Sharon:2019a}.

With the assumption of a C\,{\sc i}-to-${\rm H_2}$ abundance ratio ($X_{\rm C\,{\textsc i}}=X[{\rm C\,{\textsc i}}]/X[{\rm H_2}]=M_{\rm C\,{\textsc i}}/6M_{\rm H_2}$), one can make an alternative estimate of the total molecular gas mass. While the atomic carbon gas phase abundance does not explicitly depend on CO line detections, in practice, $X_{\rm C\,{\textsc i}}$ is estimated based on authors' CO-derived ${\rm H_2}$ gas masses and choices of $\alpha_{\rm CO}$. Therefore, the C\,{\sc i}-derived total molecular gas mass suffers from systematic uncertainties similar to those affecting the CO-derived gas mass. Choices of $X_{\rm C\,{\textsc i}}$ range from $\sim10^{-5}$--$10^{-4}$ (see \citet{Dunne:2022a} for a summary). Assuming the average  $X_{\rm C\,{\textsc i}}=6.8\times10^{-5}$ for the lensed {\it Planck}-selected DSFGs in \citet{Harrington:2021a}, we find ${M_{\rm gas}=(1.22\pm0.18)\times10^{11}(\mu/30)(6.8\times10^{-5}/X_{\rm C{\textsc i}})\,M_\sun}$. This gas mass is consistent with the (magnification-corrected) CO(1--0)-derived gas mass of $(1.6^{+0.3}_{-0.2})\times10^{11}(\alpha_{\rm CO}/4.6)\,M_\sun$ in \citetalias{Sharon:2019a} (where larger Milky Way-like $\alpha_{\rm CO}$ factors were also favored by metallicity-dependent conversion factors). Obtaining similar consistency with the CO(1--0)-derived gas mass using a lower $X_{\rm C\,{\textsc i}}$ (e.\/g.\/, $X_{\rm C\,{\textsc i}}=1.6\times10^{-5}$; \citealt{Dunne:2021a}) would require unphysically low values of $T_{ex}<0$.

We can also use our {ALMA} 1.2\,mm continuum detection to estimate the molecular gas mass via dust-to-gas scaling relations. Following the work of \citet{Scoville:2016a}, we {calculate} the gas mass using

\begin{equation}\label{eq:dust_gas_mass}
\begin{aligned}
M_{\rm mol} &= \rm 1.78\,\frac{S_{obs}}{mJy} \left(1+z\right)^{-4.8}\left( \frac{\nu_{850\mu m}}{\nu_{obs}} \right)^{3.8}\left(\frac{d_L}{Gpc}\right)^2 \\ 
&\ \rm\times\frac{6.7\times 10^{19}}{\alpha_{850\mu m}}\frac{\Gamma_0}{\Gamma_{RJ}}\times 10^{10}M_\odot.
\end{aligned}
\end{equation}

\noindent
In this equation, $M_{\rm mol}$ is the molecular gas mass, $S_{obs}$ is the observed flux density, $d_L$ is the luminosity distance, $\nu_{850\mu m}$ and $\nu_{obs}$ are the frequencies corresponding to 850\,$\mu$m and the observed frequency (1.2\,mm, in this case; i.e., 353\,GHz and 250\,GHz, respectively). 
$\alpha_{850\mu m}$ is an empirical luminosity-to-mass ratio at 850\,$\mu$m in units of $\rm ergs\,s^{-1}\,Hz^{-1}\,M_\odot^{-1}$, derived with dust emissivity index assumed to $\beta=1.8$: 

\begin{equation}\label{eq:alpha850}
\begin{aligned}
\alpha_{850\mu \rm m} = {L_{\nu_{\rm 850\mu \rm m}} \over M_{\rm mol}}  = 6.2\times 10^{19}
\end{aligned}
\end{equation}

\noindent where $\Gamma_{\rm RJ}$ accounts for the difference  between the Planck function and its Rayleigh-Jeans approximation, given by 

\begin{equation}\label{eq:GammaRJ}
\begin{aligned}
\rm {\Gamma}_{\rm RJ}(T_d,\nu_{obs}, z)= \frac{B_{\nu}}{RJ_{\nu}} =  {h \nu_{obs} (1+z) / k T_d \over{e^{h \nu_{obs} (1+z) / k T_d} -1}}.
\end{aligned}
\end{equation}

\noindent This expression scales the observed dust flux density to the observed frequency $\nu_{850}$ corresponding to 850\,$\mu m$ in the rest frame and requires an assumption of the dust temperature $T_{\rm d}$. $\rm \Gamma_0=0.71$, is the constant from calibrating $\rm\Gamma_{RJ}$ at $z=0$, $\rm T_d=25\,K$ and $\lambda=850\,\mu m$. J0901's SED fitting \citep{Saintonge:2013a} leads to $T_{\rm d}=36\,{\rm K}$. However, \citet{Scoville:2014a, Scoville:2016a} suggest a lower temperature $T_{\rm d}=25\,{\rm K}$, taking into consideration the fact that luminosity-weighted fits tend to be biased in favor of warmer ISM components. Assuming $T_{\rm d}=36\,{\rm K}$ and $25\,{\rm K}$, we derive gas masses of $M_{\rm mol}=(4.98\pm 0.51)(\mu/30)\times 10^{10}\,M_\odot$ and $(6.65\pm 0.67)(\mu/30)\times 10^{10}\,M_\odot$, respectively{, using our ALMA ${1.2\,{\rm mm}}$ data}. As found {by \citetalias{Sharon:2019a} using their lower-significance SMA} 877\,$\mu m$ detection, the continuum-derived gas mass is about 2--3 times lower than the CO-derived gas mass{; the dust temperature would have to be about ${14\,{\rm K}}$ to yield the CO(1--0)-derived gas mass}. The differences between the line-derived gas masses (both CO and \ci) and the continuum-derived gas masses may be due to a combination of factors: (1) poor constraints on $\alpha_{CO}$, (2) uncertain \ci excitation, (3) {SED-derived} dust temperatures that may be biased to hotter/more luminous regions, and{/or} (4) differential lensing. {\citetalias{Sharon:2019a} found that the dust-to-gas ratio for J0901 is consistent with expectations for sources with its redshift and metallicity, so variations in dust-to-gas ratio are unlikely to explain the discrepancy is mass estimates.} More observations and detailed modelling are required to understand the significance of these differences and potentially bring all gas masses into {better} agreement.

\subsection{CO SLED and radiative transfer modeling}\label{subsec:SLED}

With these new ALMA observations, we can begin to constrain the physical conditions of the molecular gas. Since the emission is likely optically thick, we used a large velocity gradient (LVG) model \citep[e.\/g.\/,][]{Goldreich:1974a, Scoville:1974a} to determine the CO line ratios for a range of model parameters. Since there are three model parameters in the LVG analysis (the kinetic temperature, $T_{kin}$, the molecular hydrogen density, $n_{H_2}$, and the CO column density per unit velocity gradient, $N_{\rm CO}/\Delta v$), and two measurements with which to constrain the model (CO line ratios measured relative to same transition, which we choose to be CO(1--0); see Table~\ref{tab:ratios}), the range of parameters that reasonably reproduce the observations is quite large.
We therefore compare our observed line ratios to the single-phase grid-based LVG modeling tool of \citet{Sharon:2015a}. This model assumes a spherical cloud geometry and accounts for heating from the CMB. The three parameters are sampled evenly in log-space (for the densities) or linearly (for the temperature). The sample spacing and ranges probed are the same as in \citet{Sharon:2015a}.

\begin{deluxetable*}{ccccc}
\tablewidth{0pt}
\tablecaption{Observed line luminosity ratios (in brightness temperature units) \label{tab:ratios}}
\tablehead{ \colhead{Ratio} & \colhead{North} & \colhead{South} & \colhead{West} & \colhead{Total}}
\startdata
$r_{3,1}$ & $0.74\pm0.11$ & $0.84\pm0.14$ & $0.62\pm0.11$ & $0.75\pm0.11$ \\
$r_{7,3}$ & $0.071\pm0.014$ & $0.093\pm0.019$ & $0.065\pm0.015$ & $0.078\pm0.015$ \\
$r_{7,1}$ & $0.053\pm0.011$ & $0.078\pm0.017$ & $0.040\pm0.010$ & $0.058\pm0.011$ \\
C\,{\sc i}/CO(7--6) & $1.37\pm0.24$ & $1.56\pm0.27$ & $2.04\pm0.49$ & $1.59\pm0.28$ \\
\enddata
\tablecomments{Using the naturally-weighted images only, assuming no differences in magnification factors. CO(1--0) line fluxes and the $r_{3,1}$ values are from \citetalias{Sharon:2019a}.}
\end{deluxetable*}

\begin{figure}
\centering
    \includegraphics[width=0.95\columnwidth]{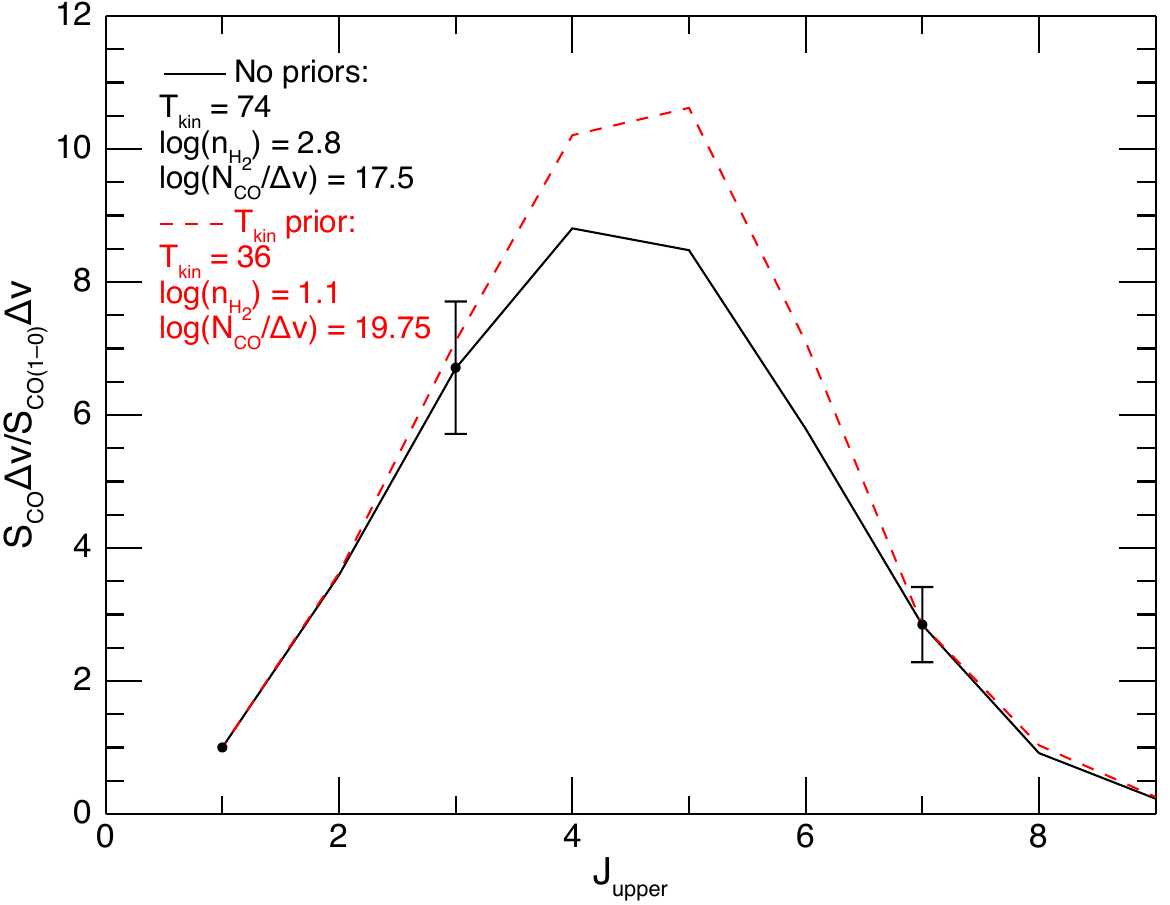}
    \caption{
    Best-fit CO SLED without any priors applied (black solid line) and with a Gaussian prior at $T_{kin}=T_{dust}$ from \citet{Saintonge:2013a} (dashed red line). Measured integrated line fluxes relative to CO(1--0) are the black points with uncertainties; the CO(1--0) line ratio with itself is one by definition, and therefore does not have an uncertainty. Best-fit parameter values are listed in the plot and given in Table~\ref{tab:lvg_params} (with units).
}
    \label{fig:COSLED}
\end{figure}

We use both minimum-$\chi^2$ and Bayesian analyses to evaluate how well each combination of model parameters reproduces our observations. Since the probabilities are related to $\chi^2$ \citep{Ward:2002a,Sharon:2015a} the best-fit models are the same when uniform priors are applied (in log or linear space, for the densities and temperature, respectively; $T_{kin}<T_{CMB}$ at the redshift of J0901 is excluded). We also test an additional prior, setting the kinetic temperature to the IR SED-determined dust temperature and uncertainty from \citet{Saintonge:2013a} ($T_{kin}=T_{dust}=36\pm1\,{\rm K}$), since gas and dust may be coupled, although LVG models do not assume local thermodynamic equilibrium. We {do not use the} \ci {line to similarly break model degeneracies since we do not have measurements of the second} \cishort {line to constraint the gas excitation temperature (which assumes local thermodynamic equilibrium, and that may not be a good assumption, e.\/g.\/, \citealt{Papadopoulos:2022a}). We} list the best-fit values with and without the $T_{kin}=T_{dust}$ prior in Table~\ref{tab:lvg_params} and compare the resulting SLEDs to the measured line ratios in Fig.~\ref{fig:COSLED}. 

\begin{deluxetable}{ccc}
\tablewidth{0pt}
\tablecaption{Best-fit LVG model parameters \label{tab:lvg_params}}
\tablehead{ \colhead{Parameter} & \colhead{Uniform priors} & \colhead{$T_{kin}=T_{dust}$ prior}}
\startdata
$T_{kin}\,{\rm (K)}$ & $74$ & $36$ \\
${\rm log}(n_{\rm H_2}/{\rm cm^{-3}})$ & $2.8$ & $1.1$ \\
${\rm log}((N_{\rm CO}/\Delta v)/({\rm cm^{-2}\,km^{-1}\,s}))$ & $17.5$ & $19.75$ \\
\enddata
\tablecomments{Due to the degeneracies in the LVG model, we do not estimate uncertainties on the best-fit model parameters (the degenerate regions in parameter space are highly non-Gaussian).}
\end{deluxetable}

As expected given the relatively small number of measurements compared to the model parameters, both best-fit models reproduce the line ratios well, with only slight differences in predicted line ratios near the peak of the CO SLED. The best-fit models' line emission is optically thick for the measured lines under consideration here. Due to the degeneracies in the LVG model, it is difficult to estimate uncertainties on the best-fit model parameters {(the degenerate regions in parameter space are highly non-Gaussian, and therefore the marginalized probability distributions do not generally correspond well} to the best-fit {parameters)}. While the flat prior best-fit model prefers kinetic temperatures $\sim 2\times$ larger than $T_{dust}$ as well as higher densities ($\log(n_{\rm H_2})=2.8$ vs. 1.1), the degeneracies are so large that it's unlikely these differences are significant. However, the best-fit density when using the $T_{kin}$ prior is sufficiently low that it is unlikely for such gas to be CO-bright \citep[e.\/g.\/,][]{Grenier:2005a}, and J0901 does not appear to have an unusually large dust-to-gas ratio shielding the CO \citep{Saintonge:2013a}. It is therefore unlikely that the $T_{kin}$ prior is accurately capturing the molecular gas physical conditions. Additional line measurements are necessary to better constrain the molecular gas physical conditions.

The observed CO(7--6)/CO(1--0) line ratio (in units of brightness temperature) of $r_{7,1}=0.058\pm0.011$ is more similar to that of local Main Sequence galaxies than, for example, the lensed {\it Planck}-selected DSFGs in \citet{Harrington:2021a}, despite the fact that J0901 has a large dust mass/luminosity. {The measured} C\,{\sc i}/CO(7--6) line ratio (in brightness temperature units) is {also} similar to that of local Main Sequence galaxies, but at the upper edge of the distribution for the lensed DSFGs selected from the South Pole Telescope \citep{Gururajan:2023a}. {However, given the degeneracies/uncertainties on the best-fit model parameters, and the differences in methodologies chosen to fit the LVG models, we cannot make robust comparisons between the physical conditions in J0901 and the latter sample}.

\subsection{Schmidt-Kennicutt relation}\label{subsec:skrelation}

Few high-redshift galaxies have sufficiently high-resolution observations to enable comparisons of gas mass and SFR surface densities on a pixel-by-pixel basis (as is done for low-redshift galaxies) in a Schmidt-Kennicutt relation \citep[e.\/g.\/,][]{Sharon:2013a, Genzel:2013a, Rawle:2014a}. In the case of J0901, $\sim1^{\prime\prime}$ resolution observations are sufficient given its large magnification factor and constructed lens models from literature. \citetalias{Sharon:2019a} showed that the slope of the Schmidt-Kennicutt relation did not differ significantly when CO(1--0) and CO(3--2) were used as the gas mass tracers (at least for imaging at the same spatial resolution and inner $uv$ radius). However, that work only had access to resolved ${\rm H\alpha}$ maps to trace the SFR, and J0901's large $L_{\rm TIR}$ indicates that a significant fraction of young stars are heavily obscured by dust. With our new observations, we are now able to extend the comparisons between gas tracers to include the \ci and CO(7--6) lines, as well as compare the performance of ${\rm H\alpha}$ to IR and radio continuum as SFR tracers.

In order to fit the distributions of pixelized SFR and gas mass surface densities to a power-law form for the Schmidt-Kennicutt relation, we employ the methodology described in \citet{Leroy:2013a} (which was based on \citealt{Blanc:2009a}) as modified in \citetalias{Sharon:2019a}. In summary, we compare 2D histograms (in log space) of the data  {(smoothed to a common resolution) to a} model-generated histogram. To generate the model, we sample the per-pixel gas mass measurements 10,000 times (with replacement), randomly perturb the individual points by their statistical uncertainty, and randomly perturb all points by their global flux calibration uncertainty. Using the perturbed gas mass data, we then calculate the expected SFRs from 

\begin{equation}\label{eq:sk_fit}
\Sigma_{\rm SFR}=A\,\left({\frac{\Sigma_{\rm gas}}{10^3\,{\rm M_\sun\,pc^{-2}}}}\right)^n\times 10^{{\mathcal N}(0,\sigma)},
\end{equation}

\noindent
for a range of normalization factors ($A$), slopes ($n$), and Gaussian distribution scatter widths about the relation ($\sigma$; we treat the mean as zero). We then compare the models' 2D histograms to the observed histogram (also with uncertainty perturbations applied, as well as a $2\sigma$ cut on the per-pixel gas mass) to calculate a $\chi^2$ goodness-of-fit statistic for all combinations of the three model parameters. In order to find the best-fit (minimum $\chi^2$) model parameters while mitigating the effects of parameter sampling, we fit the distribution of minimum $\chi^2$ values for each parameter to a third order polynomial and find the location of the polynomial's minimum. We repeat this procedure 100 times, each time generating new statistical and flux calibration perturbations and excluding a randomly chosen pixel. The mean and standard deviation of the minima locations give us the best-fit value for the parameter and our uncertainty.

In order to ease comparisons between our gas mass and SFR tracers, we scale each map such that its total SFR matches that determined from the TIR in \citet{Saintonge:2013a} with a Kroupa IMF \citep{Kroupa:2001a} and the \citetalias{Sharon:2019a} magnification factors. This method effectively creates a global offset on the SFR surface density axis (since we probe the Schmidt-Kennicutt relation in log-space) while preserving the tracers' relative distribution, leaving differences in the best-fit slope and scatter unaffected. Since the majority of the star formation activity in J0901 appears to be obscured, this scaling also ensures that our best-fit normalization is the most accurate value we can obtain (i.\/e.\/ makes a global correction for dust extinction when using H$\alpha$ as the SFR tracer, and compensates for the ALMA continuum data sampling further down the Rayleigh-Jeans tail than would be ideal for mapping ${L_{\rm IR}}$-inferred SFRs), and therefore the most useful in comparisons to others' analyses of the Schmidt-Kennicutt relation. We similarly scale the gas masses inferred from the CO(3--2) and CO(7--6) lines by their global integrated ratio with the CO(1--0) line (for the \ci line, we use the inferred gas mass derived from Equation~\ref{eq:ci_gas_mass}, since it is very close to the CO(1--0) gas mass). For each pair of gas mass and SFR tracers, we smooth the maps to match in angular resolution and re-grid the data to the same pixel sizes/locations. 

\begin{figure*}
\centering
    \includegraphics[scale=0.9]{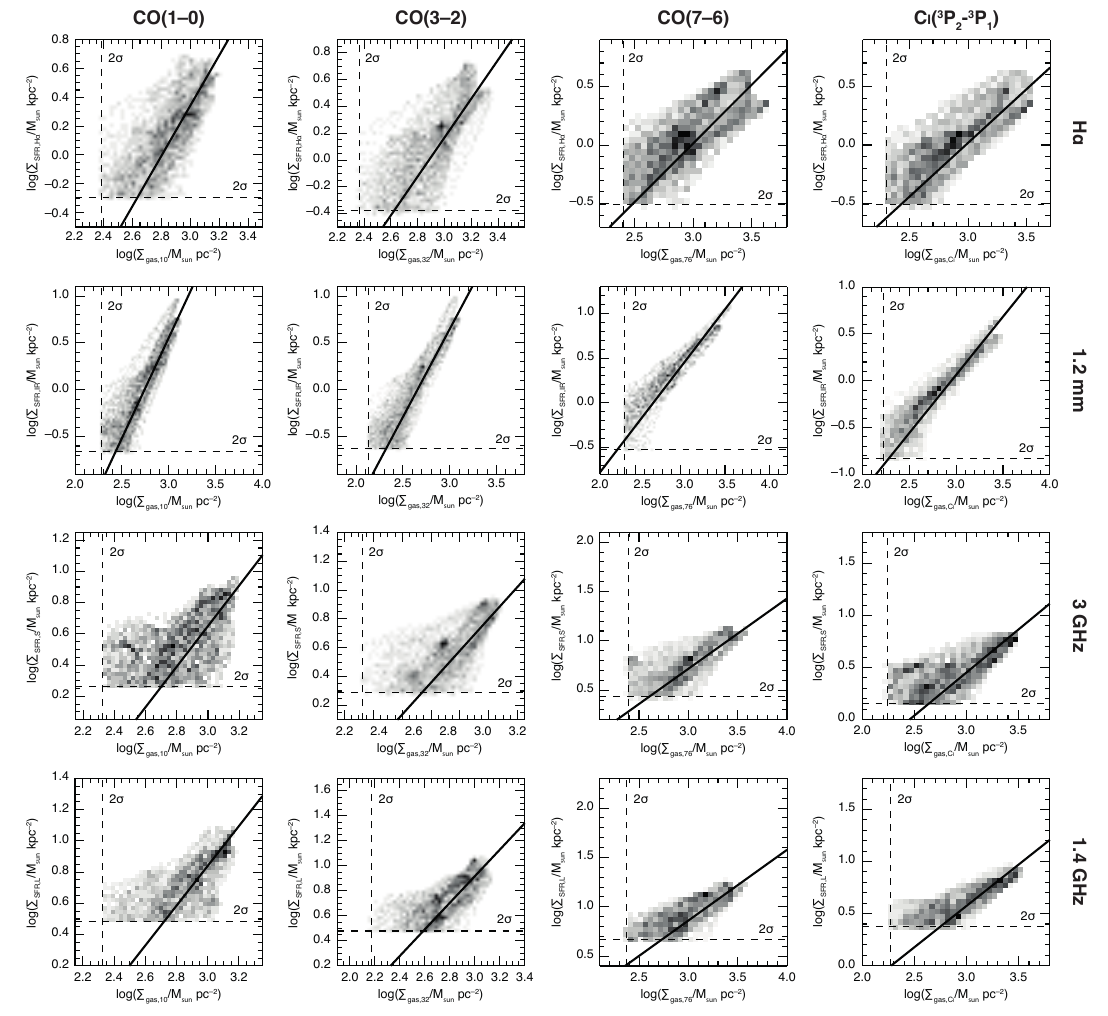}
    \caption{The resolved Schmidt-Kennicutt relation for J0901 using different gas tracers (CO(1--0): leftmost column, CO(3--2): second from left column, CO(7--6): second from right column, and \ci: rightmost column) and SFR tracers (${\rm H\alpha}$: top row, {${1.2\,{\rm mm}}$} continuum: second from top row,  {${3\,{\rm GHz}}$} radio continuum: second from bottom row, and  {${1.4\,{\rm GHz}}$} radio continuum: bottom row). Grayscale images show the number of pixels per bin in the 2D histogram in the SFR and gas mass surface density plane. The best-fit line for the Schmidt-Kennicutt relation of each panel is shown in black. The $2\,\sigma$ surface brightness cutoffs (dashed lines) are shown for clarity (although only the gas cut is applied during fitting).}
	\label{fig:sk}
\end{figure*}

We show the best-fit results in Fig.~\ref{fig:sk} and list the best-fit parameter values in Table~\ref{tab:sk}. {Since nearby pixel values are correlated due to the angular resolution of the data, we also list the number of pixels used in the fit and the corresponding number of independent beams/PSFs in Table~\ref{tab:sk}.}

For the SFR tracers, we find that the  {total SFR-scaled ALMA ${1.2\,{\rm mm}}$ continuum} maps produce the steepest slopes in the Schmidt-Kennicutt relation  {(${n=1.24}$--${2.14}$, depending on the gas mass tracer). While the ALMA continuum emission observed here is on the Rayleigh-Jeans tail of the dust emission, which is often used as a gas mass tracer, the super-unity Schmidt-Kennicutt index (${n=2.14\pm0.07}$ vs. CO(1--0))} suggests that the {observed ALMA continuum is tracing additional processes beyond just gas mass surface density. It is possible that the} regions in J0901 with the highest SFRs also have either the highest dust obscuration or the highest dust temperatures. Disentangling these two effects would require sufficiently resolved multi-band continuum observations to allow us to map the dust temperature. However, we note that the slit spectra from \citet{Hainline:2009a}, which are biased towards the central bright regions of J0901 (when de-lensed), produce ${\rm H\alpha / H\beta}$ ratios that support significant dust obscuration. {The steeper slopes found when using the ALMA continuum emission when compared to the H$\alpha$ data reinforces the importance of making spatially resolved extinction corrections when analyzing the resolved Schmidt-Kennicutt relation for high-redshift galaxies \citep[e.\/g.\/,][]{Genzel:2013a}, particularly in sources like J0901 where the majority of the star formation appears to be obscured by dust.}

\begin{deluxetable*}{llrrccc}
\tablewidth{0pt}
\tablecaption{J0901 Schmidt-Kennicutt fit parameters \label{tab:sk}}
\tablehead{ \colhead{Gas tracer} & \colhead{SFR tracer} & \colhead{$N_{pix}$} & \colhead{$N_{ind}$} & \colhead{$log(A)$} & \colhead{$n$} & \colhead{$\sigma$}}
\startdata
CO(1--0) & H$\rm \alpha$ & $4227$ & $52.6$ & $0.34\pm0.10$ & $1.74\pm0.07$ & $0.22\pm0.01$ \\ 
{} &  ${1.2\,{\rm mm}}$ & $6333$ & $42.8$ & $0.56\pm0.09$ & $2.14\pm0.07$ & $0.21\pm0.02$ \\ 
{} &  ${3\,{\rm GHz}}$ & $5725$ & $55.8$ & $0.65\pm0.09$ & $1.30\pm0.05$ & $0.20\pm0.01$ \\ 
{} &  ${1.4\,{\rm GHz}}$ & $5843$ & $53.8$ & $0.84\pm0.09$ & $1.28\pm0.05$ & $0.12\pm0.02$ \\ 
\hline
CO(3--2) & H$\rm \alpha$ & $4985$ & $52.4$ & $0.40\pm0.08$ & $1.46\pm0.04$ & $0.24\pm0.01$ \\ 
{} &  ${1.2\,{\rm mm}}$ & $6421$ & $49.4$ & $0.64\pm0.07$ & $1.88\pm0.04$ & $0.19\pm0.01$ \\ 
{} &  ${3\,{\rm GHz}}$ & $5026$ & $40.5$ & $0.81\pm0.07$ & $1.10\pm0.04$ & $0.19\pm0.02$ \\ 
{} &  ${1.4\,{\rm GHz}}$ & $4777$ & $44.8$ & $0.92\pm0.07$ & $1.06\pm0.04$ & $0.13\pm0.02$ \\ 
\hline
CO(7--6) & H$\rm \alpha$ & $2360$ & $29.4$ & $0.02\pm0.07$ & $1.00\pm0.04$ & $0.29\pm0.01$ \\ 
{} &  ${1.2\,{\rm mm}}$ & $2455$ & $27.2$ & $0.39\pm0.08$ & $1.30\pm0.09$ & $0.18\pm0.03$ \\ 
{} &  ${3\,{\rm GHz}}$ & $2599$ & $28.9$ & $0.72\pm0.05$ & $0.71\pm0.03$ & $0.18\pm0.03$ \\ 
{} &  ${1.4\,{\rm GHz}}$ & $2464$ & $27.3$ & $0.86\pm0.06$ & $0.71\pm0.02$ & $0.13\pm0.03$ \\ 
\hline
\ci & H$\rm \alpha$ & $2877$ & $34.9$ & $0.02\pm0.07$ & $0.92\pm0.04$ & $0.22\pm0.01$ \\
{} &  ${1.2\,{\rm mm}}$ & $2945$ & $28.7$ & $0.06\pm0.06$ & $1.24\pm0.04$ & $0.17\pm0.01$ \\ 
{} &  ${3\,{\rm GHz}}$ & $3392$ & $32.7$ & $0.45\pm0.06$ & $0.83\pm0.03$ & $0.19\pm0.02$ \\
{} &  ${1.4\,{\rm GHz}}$ & $3285$ & $36.0$ & $0.57\pm0.06$ & $0.79\pm0.03$ & $0.10\pm0.03$ \\ 
\hline
\enddata
\tablecomments{The table only lists the best-fit parameters for fits using all three images of J0901 combined; any image-to-image differences are discussed in the text. $N_{pix}$ lists the total number of pixels used in the fitting procedure, which are not all independent from one another due to the beam/PSF size, and $N_{ind}$ lists the number of independent resolution elements (beams/PSFs) to which those pixels correspond.}
\end{deluxetable*}

The two radio continuum maps produce the flattest slopes in the Schmidt-Kennicutt relation, despite being extinction-free tracers of the SFR. Observationally, this result is due to the fact that the  {${1.4\,{\rm GHz}}$ and ${3\,{\rm GHz}}$} maps do not seem to trace the same structures within J0901; specifically, the radio continuum for the northern arc is systematically offset radially from the lens when compared to the other data for that image\footnote{When the northern image is analyzed separately, the radio continuum emission is essentially uncorrelated with the gas tracers when plotted on the Schmidt-Kennicutt relation.}, the two peaks of emission in the southern image are more widely separated in the radio continuum than in the other data, and only the radio data shows two peaks of emission in the western image. While the radio continuum data have lower S/N than the other SFR tracers (note the $2\sigma$ cuts in Fig.\ref{fig:sk}, although those SFR cuts are not applied when fitting the Schmidt-Kennicutt relation; only gas cuts are applied), data quality is unlikely to explain all of these differences, particularly when the same effects are seen in two different bands. The radio  {spectrum} does not show strong evidence that the central AGN in J0901 is producing significant synchrotron emission. However, our radio continuum data lack the S/N necessary to map the radio spectral index and look for features such as compact steep spectrum radio emission from the AGN \citep[e.\/g.\/,][]{ODea:2021a} or localized clusters of supernova remnants associated with clumpy star formation. \citet{Thomson:2019a} find that the radio continuum emission in their sample of dusty star-forming galaxies is more extended than the dust emission, and suggest that the production of secondary low-energy cosmic ray electrons far from the nuclear starburst may be producing the size mismatch. It is possible that a less symmetric version of this effect is occurring within J0901 if extreme density inhomogeneities are preventing/favoring the spread of cosmic ray nucleons in certain directions. 

\begin{figure}
\centering
    \includegraphics[width=0.95\columnwidth]{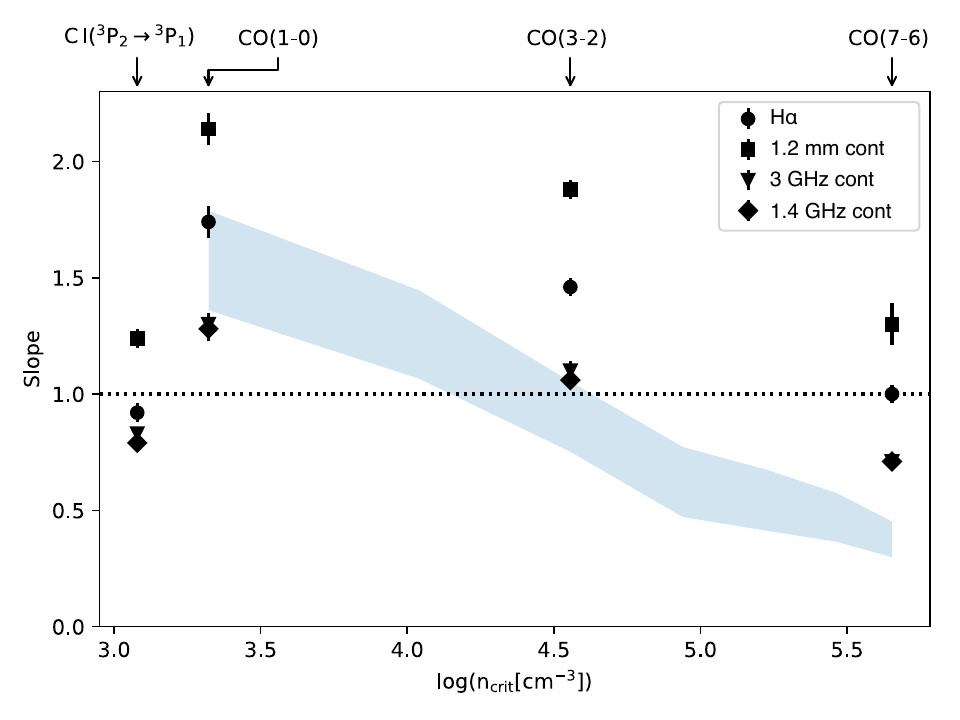}
    \caption{
    The SK relation best-fit slope in the function of lines' critical densities. The blue shaded region shows the distribution of indices found in the simulation of \cite{Narayanan:2008a}.
}
    \label{fig:ncrit}
\end{figure}

Among the CO lines being used as molecular gas tracers, we find the higher-excitation lines are associated with flatter slopes for the Schmidt-Kennicutt relation. The same trend has been seen for integrated luminosity-luminosity versions of the Schmidt-Kennicutt relation for local galaxies (e.\/g.\/, \citealt{Greve:2014a, Lu:2017a}; cf. \citealt{Liu:2015a, Kamenetzky:2016a}). Moreover, the same trend as a function of critical density was found for galaxy simulations in \citet{Narayanan:2008a, Juneau:2009a}. We plot the measured slopes as a function of gas tracer critical density, $n_{crit}$ (from \citealt{Carilli:2013a}) in Fig.~\ref{fig:ncrit}, and find that J0901 is generally offset to steeper slopes than predicted in \citet{Narayanan:2008a} (depending on the SFR tracer being used), and that the slopes decline with $n_{crit}$ more slowly than predicted, although we note that those simulations are for unresolved observations. We do note that \citetalias{Sharon:2019a} found that the difference between the slopes of the Schmidt-Kennicutt relations for the CO(1--0) and CO(3--2) lines disappeared when the data were smoothed to match spatial resolutions and clipped to the same inner $uv$-radius before imaging. Performing similar smoothing across the seven interferometric datasets analyzed here (but no $uv$ clipping) does show that more smoothing results in flatter slopes. However, the overall trend of flattening slope with increasing critical density remains (since the slopes flatten by at most $\sim10\%$ for the smoothing necessary to match all data); while the small difference between the CO(1--0) and CO(3--2) slopes falls below statistical significance for ${\rm H\alpha}$ once smoothed, it does not do so when IR is used as the SFR tracer, for example. Thus, using even mid-$J$ CO lines to measure the spatially resolved Schmidt-Kennicutt relation may alter conclusions at high redshift.

Despite yielding a total gas mass similar to that for the CO(1--0) line and having a similar critical density, the \ci line does not show a similar slope for the Schmidt-Kennicutt relation. The \ci line behaves more similarly to the close-by CO(7--6) line. The \ci line's dissimilarity to the CO(1--0) emission may be due to the former being subthermally excited \citep[e.\/g.\/,][]{Papadopoulos:2022a}, particularly if the ${\rm H_2}$ gas density is as low as implied by the LVG modeling results (Section~\ref{subsec:SLED}). In this scenario, \ci emission is weak and undetected in regions that are producing the CO(1--0) emission, leading to the observed difference in the Schmidt-Kennicutt relations. It is possible that the C\,{\sc i}\,($^3P_1\rightarrow^3P_0$) line will better match the low-$J$ CO gas tracers.

\section{Summary}\label{sec:concl}

We have presented VLA continuum {observations at observed frequencies of ${1.4\,{\rm GHz}}$ and ${3\,{\rm GHz}}$, as well as ALMA observations of the ${1.2\,{\rm mm}}$ (observed) continuum,} CO(7--6) {line,} and \ci line, for the strongly lensed galaxy J0901 at $z=2.26$. We summarise our  conclusions as follows.

\begin{itemize}
    \item We {resolved} the radio continuum emission from J0901 at $\sim 1.\arcsec1$ resolution at both 1.4 and $3\,{\rm GHz}$ (observed frame) using the VLA. All three images are clearly detected. The integrated flux densities measured at the two bands are ${{S_{\rm 1.4\,GHz}=\rm 503\pm 42\,\mu Jy}}$ and ${{S_{\rm 3\,GHz}=\rm 328\pm 28\,\mu Jy}}$.

    \item We resolve J0901 in the CO(7--6) and \ci lines, as well as the  {${1.2\,{\rm mm}}$ (observed frame)} continuum, using ALMA, with a resolution of $1.\arcsec7\times 1.\arcsec1$. We measure integrated line fluxes of $8.4\pm1.4\,{\rm Jy\,km\,s^{-1}}$ for CO(7--6) and $13.4\pm2.0\,{\rm Jy\,km\,s^{-1}}$ for \ci, and a  continuum flux density of $\rm 14\pm1.4\,mJy$.

    \item Combining the integrated flux densities measured from the 1.4\,GHz, 3\,GHz, and previously obtained $35\,{\rm GHz}$ continuum detections, we decompose the radio emission into FF emission and non-thermal components. We run two series of MCMC simulations, with flat and Gaussian priors on the non-thermal spectral index. The resulting best-fit  {spectrum} indicates that the higher frequency radio continuum is dominated by FF rather than non-thermal emission, with  ${f_{th}=0.31_{-0.19}^{+0.31}}$ at rest-frame 1.4\,GHz (for the Gaussian prior). The thermal fraction is  {high relative to the typical value of ${\sim0.1}$ found for} star-forming galaxies {at both low and high redshifts \citep[e.\/g.\/,][]{Condon:1992a, Algera:2021a}, but is technically consistent due to the large uncertainty. The large uncertainties in the thermal fractions are driven by poor constraints on} the non-thermal spectral index  {(even when using the Gaussian prior), with ${{\rm\alpha^{NT}=-1.00^{+0.44}_{-0.45}}}$}.

    \item We measure the IRRC $q_{\rm TIR}=2.65_{-0.31}^{+0.24}$ using the 1.4\,GHz (rest-frame) emission from the fit to the radio spectrum (for the Gaussian prior). This value is consistent with expectations for star-forming galaxies at high redshift and in the Local Universe. Coupled with the spectral index constraints, these results indicate that the weak AGN previously identified in J0901 is not radio-loud.

    \item We use the best-fit radio {spectrum} (for the  Gaussian prior) to calculate the SFR for J0901.  Using the combination of the FF and synchrotron emission directly yields a SFR of ${\rm 250_{-150}^{+470}\,M_\odot\,yr^{-1}}$, which is consistent with the TIR-inferred SFR given the considerable uncertainties in the radio {spectrum} decomposition.

    \item We use the ALMA detections of the dust continuum and \ci line to estimate the gas mass in J0901. The \ci-derived gas mass, assuming an average C\,{\textsc i}-to-${\rm H_2}$ abundance ratio $X_{\rm C\,{\textsc i}}=6.8\times10^{-5}$, is $M_{\rm gas}=(1.3\pm0.2)\times10^{11}(\mu/30)(6.8\times10^{-5}/X_{\rm C{\textsc i}})\,M_\sun$. This value is consistent with the CO(1--0)-derived gas mass in \citetalias{Sharon:2019a}. Using the $1.2\,{\rm mm}$ continuum and relations from \citet{Scoville:2016a}, we find a gas mass of $(6.65\pm 0.67)\times 10^{10}\,M_\odot$ assuming a dust temperature $T_{\rm d}=25\,{\rm K}$. This mass is $\sim2$--$3\times$ lower than the CO-derived gas mass, and suggests that lower values of the CO-to-H$_2$ conversion factor (more in line with those appropriate for local U/LIRGs) may be more appropriate for J0901. Differential lensing of line and continuum emission may also affect the consistency of our inferred gas masses.

    \item  {The observed CO(7--6)/CO(1--0) and C\,{\sc i}/CO(7--6) line ratios are more similar to those of local Main Sequence galaxies than those of high-redshift DSFGs. Fitting LVG models to the} independent CO line ratios  {results in a highly degenerate parameter space with poor constraints on} the gas physical conditions.

    \item  {We compare how the Schmidt-Kennicutt relation changes with choice of gas tracer} (CO(1--0), CO(3--2), CO(7--6), and \ci) and  {SFR tracer} (${\rm H\alpha}$, ${1.2\,{\rm mm}}$ IR continuum,  {${3\,{\rm GHz}}$, ${1.4\,{\rm GHz}}$} radio continuum). We find that the ${1.2\,{\rm mm}}$ continuum produces the steepest slopes ($n\sim 1.2$--2) regardless of gas tracer, while the radio continuum maps produce the flattest slopes ($n\sim0.7$-1.3). The IR results suggest the regions with the highest SFRs have high dust obscuration and/or dust temperatures. The flatter Schmidt-Kennicutt slopes for the radio continuum are likely driven by the poorer spatial correspondence between the radio emission and other gas and SFR tracers{.}

    \item The slope of the resolved Schmidt-Kennicutt relation decreases with the CO line critical density, as seen in unresolved samples of galaxies. However, when compared across SFR tracers, we find significant flattening for even the CO(3--2) line, indicating that such mid-$J$ CO lines may produce inconsistent results for spatially resolved mapping of the Schmidt-Kennicutt relation. The \ci line also produces flatter slopes, most similar to those for the CO(7--6) line, potentially due to the low density of the gas and sub-thermal excitation. 
\end{itemize}

\begin{acknowledgments}
We thank the anonymous referee for the helpful comments and suggestions, which have improved our paper. This paper makes use of the following ALMA data: ADS/JAO.ALMA\#2013.1.00952.S. ALMA is a partnership of ESO (representing its member states), NSF (USA) and NINS (Japan), together with NRC (Canada), MOST and ASIAA (Taiwan), and KASI (Republic of Korea), in cooperation with the Republic of Chile. The Joint ALMA Observatory is operated by ESO, AUI/NRAO and NAOJ. The National Radio Astronomy Observatory is a facility of the National Science Foundation operated under cooperative agreement by Associated Universities, Inc. 
This research is supported by the Ministry of Education, Singapore, under its Academic Research Fund Tier 2 program (T2EP50121-0016).
HSBA acknowledges support from the NAOJ ALMA Scientific Research Grant Code 2021-19A.
AJB acknowledges support from the National Science Foundation through grant AST-1716585 and from the Radcliffe Institute for Advanced Study at Harvard University.

\end{acknowledgments}

\bibliographystyle{apj}

\end{document}